\documentclass[11pt,a4paper]{article}
\usepackage{jinstpub}

\usepackage[utf8]{inputenc} % allow utf-8 input
\usepackage[T1]{fontenc}    % use 8-bit T1 fonts
\usepackage{hyperref}       % hyperlinks
\usepackage{url}            % simple URL typesetting
\usepackage{booktabs}       % professional-quality tables
\usepackage{amsfonts}       % blackboard math symbols
\usepackage{nicefrac}       % compact symbols for 1/2, etc.
\usepackage{microtype}      % microtypography
\usepackage{lipsum}
\usepackage{graphicx}
\usepackage{ulem}
\usepackage{float}
\usepackage{amsmath}

\title{Large scale characterization and calibration strategy of a SiPM-based camera for gamma-ray astronomy}

\author[a,1]{C.~Alispach\note{Corresponding author.},}
\author[e]{J.~Borkowski,}
\author[a]{F.~R.~Cadoux,}
\author[a]{N.~De~Angelis,}
\author[a]{D.~Della~Volpe,}
\author[a]{Y.~Favre,}
\author[a]{M.~Heller,}
\author[d,f]{J.~Jury{\v{s}}ek,}
\author[b]{E.~Lyard,}
\author[d]{D.~Mandat,}
\author[a]{L.~David~M.~Miranda,}
\author[a]{T.~Montaruli,}
\author[a]{A.~Nagai,}
\author[c]{D.~Neise,}
\author[a]{T.~R.~Njoh~Ekoume,}
\author[d]{M.~Pech,}
\author[h]{P.~Rajda,}
\author[a]{Y.~Renier,}
\author[b]{V.~Sliusar,}
\author[b]{R.~Walter}
\author[g]{and K.~Zi{\c{e}}tara}
\affiliation[a]{Universit{\'e} de Gen{\`e}ve, DPNC,\\Quai Ernest Ansermet 24, CH-1211, Gen{\`e}ve, Switzerland}
\affiliation[b]{Universit{\'e} de Gen{\`e}ve, D{\'e}partement d'Astronomie,\\Chemin d'Ecogia 16, CH-1290, Versoix, Switzerland}
\affiliation[c]{ETH Zurich, Institute for Particle Physics and Astrophysics,\\Otto-Stern-Weg 5, CH-8093, Zurich, Switzerland}
\affiliation[d]{FZU - Institute of Physics of the Czech Academy of Sciences,\\17. listopadu 50, Olomouc {\&} Na Slovance 2, Prague, Czech Republic}
\affiliation[e]{Nicolaus Copernicus Astronomical Center, Polish Academy of Sciences,\\ul. Bartycka 18, 00-716, Warsaw, Poland}
\affiliation[f]{Palacky University Olomouc, Faculty of Science, RCPTM,\\17. listopadu 50, Olomouc, Czech Republic}
\affiliation[g]{Astronomical Observatory, Jagiellonian University,\\ul. Orla 171, 30-244, Krak{\'o}w, Poland}
\affiliation[h]{AGH University of Science and Technology,\\ al.Mickiewicza 30, 30-059 Krak{\'o}w, Poland}
\emailAdd{cyril.alispach@unige.ch}

\abstract{The SST-1M is a 4-m diameter mirror Davies-Cotton gamma-ray telescope. It has been designed to cover the energy range above $\sim 500$~GeV and to be part of an array of telescopes separated by $\sim 150-200$~m. 
Its innovative camera is featuring large area hexagonal silicon photo-multipliers as photon detectors and a fully digital trigger and readout system.
Here, the strategy and the methods for its calibration are presented, together with the obtained results. 
In particular, the off and on-site calibration strategies are demonstrated on the first camera prototype. 
The performances of the camera in terms of charge and time resolution are described.}
\keywords{gamma-ray astronomy, silicon photomultiplier, Cherenkov camera}
\arxivnumber{2008.04716} % only if you have one

\begin{document}
\maketitle

\section{Introduction}\label{sec:intro}
Silicon Photo-Multipliers (SiPMs) are becoming a standard for low light level detection in nuclear and particle physics experiments. 
Presently, the majority of Cherenkov cameras used in gamma-ray astronomy on imaging air Cherenkov Telescopes (IACTs) are equipped with photo-multiplier tubes (PMTs). Their operation in the presence of moonlight is possible, but it requires either lowering the high-voltage applied to the PMTs or the use of UV-pass filters~\cite{VERITAS-moon-light, MAGIC-moon-light}. The FACT telescope~\cite{FACT-telescope} has shown that SiPMs can replace PMTs for Cherenkov cameras allowing to observe during strong moonlight and even with the Moon in the field of view (FoV), therefore, increasing the duty cycle~\cite{FACT-moon-light}, i.e. the observation time. 
Over the years, SiPMs have shown to have many advantages when compared to PMTs. As example their insensitivity to magnetic fields, the absence of ageing or the high level of reproducibility, being solid state devices, and their progressively lower cost. In terms of performance, SiPMs show consistent improvements of the photo-detection efficiency towards ultra-violet light, while reducing the correlated noise probability and dark noise rate. 
For these reasons, SiPMs are taking over PMTs in many fields of particle and nuclear physics. 
The new generation of gamma-ray observatory, the Cherenkov Telescope Array~\cite{CTA-concept} (CTA), adopts SiPMs for the small size telescope (SST) cameras~\cite{CTA-design}. As a matter of fact, the SST-1M (single mirror small size telescope)~\cite{SST-1M-TDR} has been originally designed to meet the requirements of CTA array of small size telescopes in the energy region above 1~TeV. Also the LHAASO experiment adopted SiPM for the cameras of the fluorescence/Cherenkov array of telescopes~\cite{LHASSO-SiPM} and the TAIGA experiment is also considering them for their IACTs~\cite{TAIGA-TDR}.

The SST-1M camera is described in detail elsewhere~\cite{SST-1M-camera}, and here only the major features are recollected. 
The photo-detection plane (PDP) is composed of 1296 pixels, distributed in 108 modules, made of 12  hexagonal SiPMs S10943(X) from Hamamatsu (described in~\cite{SST-1M-SiPM}) each combined with a hollow light-concentrator (described in~\cite{SST-1M-light-cones}). The cones reflect the light on the sensors, so they are optimized for almost horizontal reflectance on their surface in the blue-UV. Details on the front-end electronic can be found in~\cite{SST-1M-front-end}. 
The PDP is protected by an entrance window, a 3.3~mm-thick borofloat glass, covered with anti-reflective and low-pass filter coatings. The light above 540~nm of wavelength is cut-out to reduce the effect of the night-sky background (NSB), thus enhancing the signal-to-noise ratio. 
It is important to characterize how these elements affect the conversion factor between an impinging photon on the camera window and the number of detected photo-electrons (p.e.) by the digitizing electronics, meaning how to convert the raw data into calibrated data. 
The trigger and readout system of the camera is a fully digital system, therefore named DigiCam~\cite{SST-1M-Digicam-ICRC15}. 
In DigiCam, the raw data analog signal is constantly digitized at 250 million samples per second and the trigger decision is elaborated in an FPGA chip, where these digital signals are fed. The same data are also packed and sent to the the readout, ensuring a one-to-one correspondence between the readout and the trigger path.

This work defines the methods to measure the relevant calibration parameters in the laboratory and monitoring during operation on the field of a SiPM-based camera for gamma-ray astronomy. 
Here is described the precision at which the calibration parameters are measured together with the methods and the calibration devices used for their determination.
Through the measurement of these parameters, the camera performance can be fully assessed.
The delivery schedule for the CTA southern observatory imposes a production and delivery of two telescopes per month for a total of up to 70 SSTs. The CTA project imposes strict requirements on the performances of its instruments. These are verified during the production phase. Therefore, emphasis has been put on the automation of the calibration (see section~\ref{sec:off-site}) and validation (see section~\ref{sec:key_perf}) processes for large scale production of cameras. 

In section~\ref{sec:strategy} we present the reconstruction of the arrival time and the number of photons from the waveform signals. 
In section~\ref{sec:off-site}, we describe the mandatory characterization, in the laboratory, of the different camera elements and of the full assembled camera. The characterization is performed using a likelihood maximization of the multiple photo-electron spectra of each pixel and for various light intensity. Compared to previous work, here the likelihood includes the optical cross-talk as modeled in~\cite{SiPM-Vinogradov}. It also combines the multiple photo-electron spectra of various light intensity to achieve better precision.  
The characterization of the camera elements is essential to understand the camera performance and define the methods and strategy to calibrate the signals at the site.
For the calibration in the laboratory, the camera test setup (CTS) is a key element, therefore a brief description is also given there.
The precision achieved for the measurement of the critical parameters is presented.
Finally, the optical efficiencies of the camera elements are presented. In particular, the optical efficiency of the filtering window has been determined as function of wavelength and on its entire surface.
In section~\ref{sec:key_perf}, we present two of the most important parameters to assess the camera performance, the charge and time resolution of the camera as a function of the impinging light intensity for different NSB levels.
Finally, in section~\ref{sec:on-site}, we describe the calibration and monitoring strategy envisaged to be performed on-site. The on-site calibration provides information about the trigger stability, the aging of the PDP components and the correction of the sensor bias voltage drop induced by the NSB.

\section{Signal reconstruction}\label{sec:strategy}

The main scope of the calibration of a SiPM based gamma-ray camera is to convert raw data expressed in Least Significant Bit (LSB) units into a number of photons per pixel and a respective arrival time. 
Ultimately, the number of photons and the reconstructed arrival time profile of an extensive air-shower event are used to determine the energy and direction of the primary particle.
The precision at which they are measured is linked to the energy and angular resolution of the detector, respectively. The determination of the characterizing parameters of the camera is also relevant to accurately simulate the response of the camera to Cherenkov light.
This is particularly relevant to estimate the performance of the telescope and properly infer the fluxes of gamma-ray sources. It is also relevant when energy and direction of the primary particle are reconstructed with machine learning algorithms, trained on Monte Carlo simulated images of extensive air-showers.  

\subsection{Reconstruction of the number of photons}\label{sec:photon-reco}

\begin{figure}[htbp] 
\centering 
\includegraphics[width=0.95\textwidth]{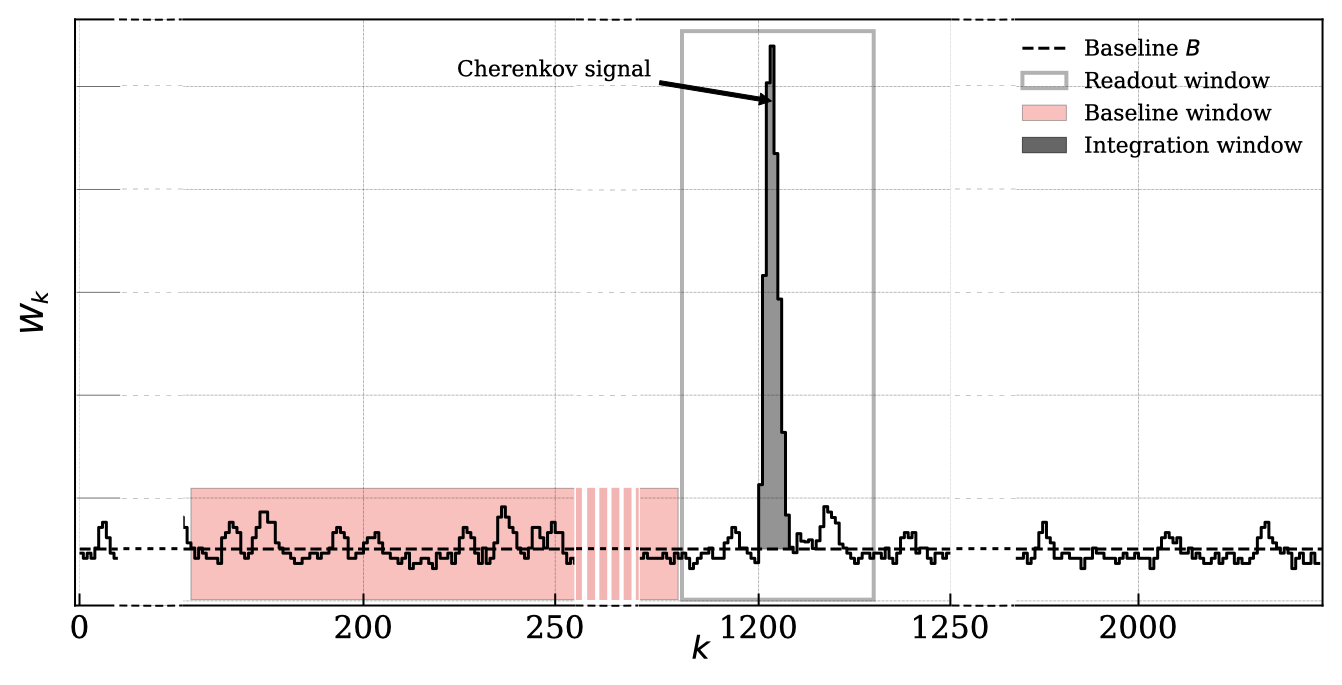}\hfill
\caption{Waveform scheme as seen in the ring buffer of DigiCam FPGA. 
The readout window contains between 50 and 91 samples depending on the used configuration. The charge is computed as the sum of the samples around the Cherenkov signal after the baseline subtraction (as in equation~\ref{eq:pe_conversion}). To be able to represent the full buffer size the waveform is truncated.
}
\label{fig:waveform_scheme}
\end{figure}

The DigiCam FPGA features a ring buffer capable of storing  up to 2048 samples at $f=250$~MHz. Each sample is a 12-bit integer marked as $W_k$, with $k \in [0, 2047]$. 
When a trigger (internal or external) is satisfied, the raw signal waveform per pixel, which is composed by an ensemble of ring-buffer samples $W_k$, is acquired in a readout window, as it is shown in figure~\ref{fig:waveform_scheme}. 
To reconstruct the number of photons, the raw waveform needs to be converted into a photon signal.
The first step is to extract the number of photo-electrons $N_{\textrm{p.e.}}$ from the waveform of the pixel. The signal baseline is mainly due to the average uncorrelated noise, such as the dark noise or NSB. Therefore, it needs to be subtracted from each waveform, to extract the signal contribution.
The baseline $B$ is computed by the DigiCam FPGA for each pixel as the average of the 1024 samples ($4.096~\mu$s) preceding the readout window of the triggered event (see equation \ref{eq:baseline_comp}).

\begin{eqnarray}
    B & = & \frac{1}{1024} \sum_{k=i}^{i+1023}W_{k} \label{eq:baseline_comp} ; \\
    N_{\textrm{p.e.}} & = & \frac{1 - \mu_{\textrm{XT}}}{G} \sum_{k=j}^{j + f\Delta t} (W_k - B), \label{eq:pe_conversion}
\end{eqnarray}

where $i$ is an adjustable parameters of the DigiCam. It allows to adjust from which ring buffer sample $k$ the baseline is calculated. This parameter is set such that the baseline is computed before the start of the readout window.

The baseline-subtracted waveforms are then integrated within the readout window over a duration $\Delta t$ from the sample $j$, where $j$ is the start of the integration window. The resulting value is converted into photo-electrons using the gain, $G$, and corrected for the average excess of photo-electrons induced by optical cross-talk $\mu_{\textrm{XT}}$, as in equation~\ref{eq:pe_conversion}.\footnote{$\mu_{\textrm{XT}}$ represents the average number of total cross-talk avalanches initiated by a primary avalanche, while  $\sum_{i=0}^{\infty} \mu_{\textrm{XT}}^i = \frac{1}{1-\mu_{\textrm{XT}}}$ is the total number of discharged micro-cells resulting from a primary avalanche.}
In order to maximize the signal-to-noise ratio, an optimal integration window of $\Delta t = 28$~ns was found.

Finally, the number of photons arriving at the camera photo-detection plane $N_{\gamma}$ is obtained by correcting also for the window transmittance $\eta_{\textrm{window}}$, the cone collection efficiency $\eta_{\textrm{cones}}$ (this includes the coating reflectance and geometrical fill factor) and the photo-detection efficiency (PDE) of the SiPM sensor, $\eta_{\textrm{SiPM}}$:  
\begin{equation}
    N_{\gamma}  = \frac{N_{\textrm{p.e.}}}{\eta_{\textrm{window}}\times\eta_{\textrm{cones}}\times \eta_{\textrm{SiPM}}}.
    \label{eq:gamma_conversion}
\end{equation}
Since the PDP efficiency depends on the photon's wavelength, we refer to the efficiency of the various camera components $\eta_{\textrm{window}}$, $\eta_{\textrm{cones}}$ and $\eta_{\textrm{SiPM}}$ as their wavelength dependent efficiencies folded with the Cherenkov signal light spectrum.\footnote{The efficiency of an optical component is $\eta = \frac{\int_0^{\infty}g(\lambda)\frac{dN}{d\lambda}d\lambda}{\int_0^{\infty}\frac{dN}{d\lambda}d\lambda}$, where $g(\lambda)$ is the transfer function of the optical component and $\frac{dN}{d\lambda}$ is the number of photons emitted by the source per unit wavelength $\lambda$.}

Among the calibration parameters defined above, the gain $G$, the baseline $B$ and the optical cross-talk $\mu_{\textrm{XT}}$ are measured for each camera pixel. Therefore the number of photo-electrons $N_{\textrm{p.e.}}$ can be computed per pixel. 
On the contrary, the overall photo-detection plane efficiency $\eta_{\textrm{total}}$ cannot be calculated for each pixel individually.
As a matter of fact, the determination of $N_{\gamma}$ requires to measure individually the SiPM photo-detection efficiency and the cone collection efficiency, which are quite time consuming procedures. However the transmittance of the camera window is determined on its entire surface (see section~\ref{sec:window}). On the other hand, the relative pixel-to-pixel PDE can be determined using a diffuse light source, with a known irradiance profile (see section~{\ref{sec:flasher}}), by comparing the reconstructed number of photo-electrons in each pixel. This relative efficiency is converted in an absolute efficiency, using  a set of 12 reference-calibrated pixels (corresponding to a PDP module), whose optical efficiencies have been precisely measured.

It is important to stress that the SiPM gain, and therefore the overall gain $G$, the optical cross-talk $\mu_{\textrm{XT}}$ and the photo-detection efficiency of the SiPM, $\eta_{\textrm{SiPM}}$, depend on the operation voltage of the SiPM.\footnote{The overall gain $G$ corresponds to the average signal charge produced by the full readout chain (SiPM, front-end pre-amplifying electronics, and back-end digitizing electronics) for one photo-electron.}
To prevent a high current to flow through the SiPM, which increases its temperature and in the long term might damage it, it is necessary to have a resistor at its bias stage which acts as a current limiter. The steady current produced by NSB light induces a voltage drop across the resistor, which reduces the operational voltage of the SiPM~\cite{SST-1M-voltage-drop}. 
As a consequence, these parameters vary for different NSB levels. This voltage drop effect has been both modeled and measured~\cite{SiPM-NSB} and it is corrected for in this analysis. 
For simplicity, the reported values in this paper for $G$ and $\mu_{\textrm{XT}}$ are given in the absence of any background light. Nonetheless, the necessary corrections are applied in the presence of background light.

\subsection{Reconstruction of the arrival time of photon light pulse}\label{sec:time-reco}

For atmospheric showers, the typical time spread of the Cherenkov signal within a single pixel (on average $\sim 1.7$~ns for diffuse cosmic ray protons with a spectrum of $E^{-2.7}$) is much shorter than the duration of the SiPM response signal (full width at half maximum $\sim 20$~ns). Thus, the Cherenkov flashes collected in each pixel can be viewed as an instantaneous flash of light. Moreover, the flash of light produces a waveform with the same shape as a typical single photo-electron pulse, but with an amplitude and integral proportional to the number of photo-electrons producing it (in the linear range of the amplification chain). 
As a consequence, a typical normalized SiPM pulse or pulse template, $h(t)$, fitted to the waveform is used to determine the arrival time of the photons. 
The procedure to derive the pulse template from measurements is illustrated in section~\ref{sec:pulse_template}.

 To extract the photon arrival time $t'$, the waveforms (composed of $N_{\textrm{samples}}$ samples) and the template are normalized to their integrals around their maximum. One defines $k_{\textrm{max}} = \underset{k}{\textrm{argmax}}~W_k$ as the sample where the waveform is maximum and $\hat{k}_{\textrm{max}} = \underset{k}{\textrm{argmax}}~h(k/f+dt)$ as the sample of the maximum of the pulse template, sampled with the time offset $dt$ with respect to the $k^{th}$ sample (sampled at $t_k = k/f$, where $f$ is the sampling frequency of FADCs). The normalization coefficients $A_W$ and $A_h(dt)$ are obtained with the sum over eight consecutive samples (three before the maximum and four after) for the waveform, once the baseline has been subtracted, as
\begin{equation}
    A_W= \sum_{k=-3}^4 \left(W_{k + k_{\textrm{max}}} -B\right)
\end{equation}
 and for the template as
\begin{equation}
    A_{h}(dt)= \sum_{k=-3}^4 h(t_k +\hat{k}_{\textrm{max}}/f + dt) \, .
\end{equation}
The residual $r_k(dt)$ of the comparison of the normalized waveform with the normalized template is:
\begin{equation}
r_k(dt)=\frac{W_k - B}{A_W} - \frac{h(t_k+dt)}{A_h(dt)}\, ,
\end{equation}
while the estimated error is the normalized combination of the electronic noise $\sigma_e$ and the uncertainty on the template $\sigma_{h(t)}$ with:
\begin{equation}
\sigma_k(dt) = \sqrt{\left(\frac{\sigma_e}{A_W}\right)^2 + \left(\frac{\sigma_{h(t_k+dt)}}{A_h(dt)}\right)^2} \,.
\end{equation}
The $\chi^2(dt)$ obtained comparing the normalized pulse template and the normalized waveform is computed for different offsets $dt$ of the template in time according to :
\begin{equation}
    \chi^2(dt) = \frac{1}{N_{\textrm{samples}}} \sum_{k=1}^{N_{\textrm{samples}}} \frac{r_k^2(dt)}{\sigma_k^2(dt)},
\end{equation}
and the photon arrival time $t'$, defined as the time where the fitted amplitude is maximum, is calculated as:
\begin{equation}\label{eq:chi2_timing}
t'= \frac{k_{\textrm{max}}}{f} -\underset{dt}{\textrm{argmin}}~\chi^2(dt) \, .
\end{equation}

\section{Off-site calibration strategy}\label{sec:off-site}

\subsection{Photo-sensor and electronics properties}
\subsubsection{The Camera Test Setup and its LED calibration}
\label{sec:ls_cts}

\begin{figure}[htbp]
\centering
\includegraphics[width=0.32\textwidth]{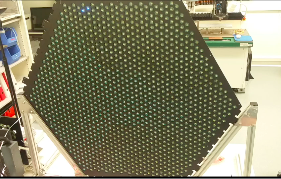}\hfill
\includegraphics[width=0.27\textwidth]{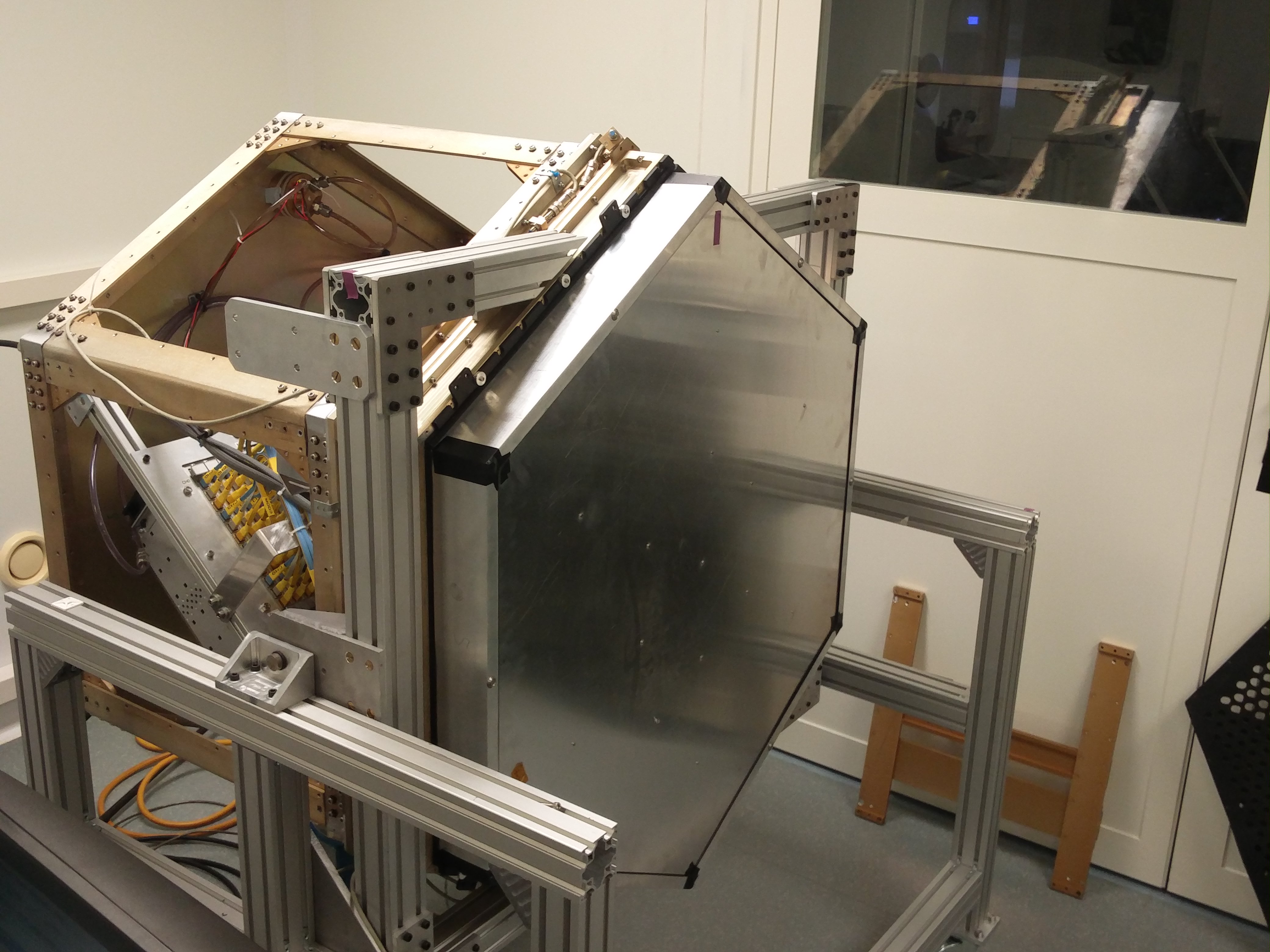} \hfill
\includegraphics[width=0.37\textwidth]{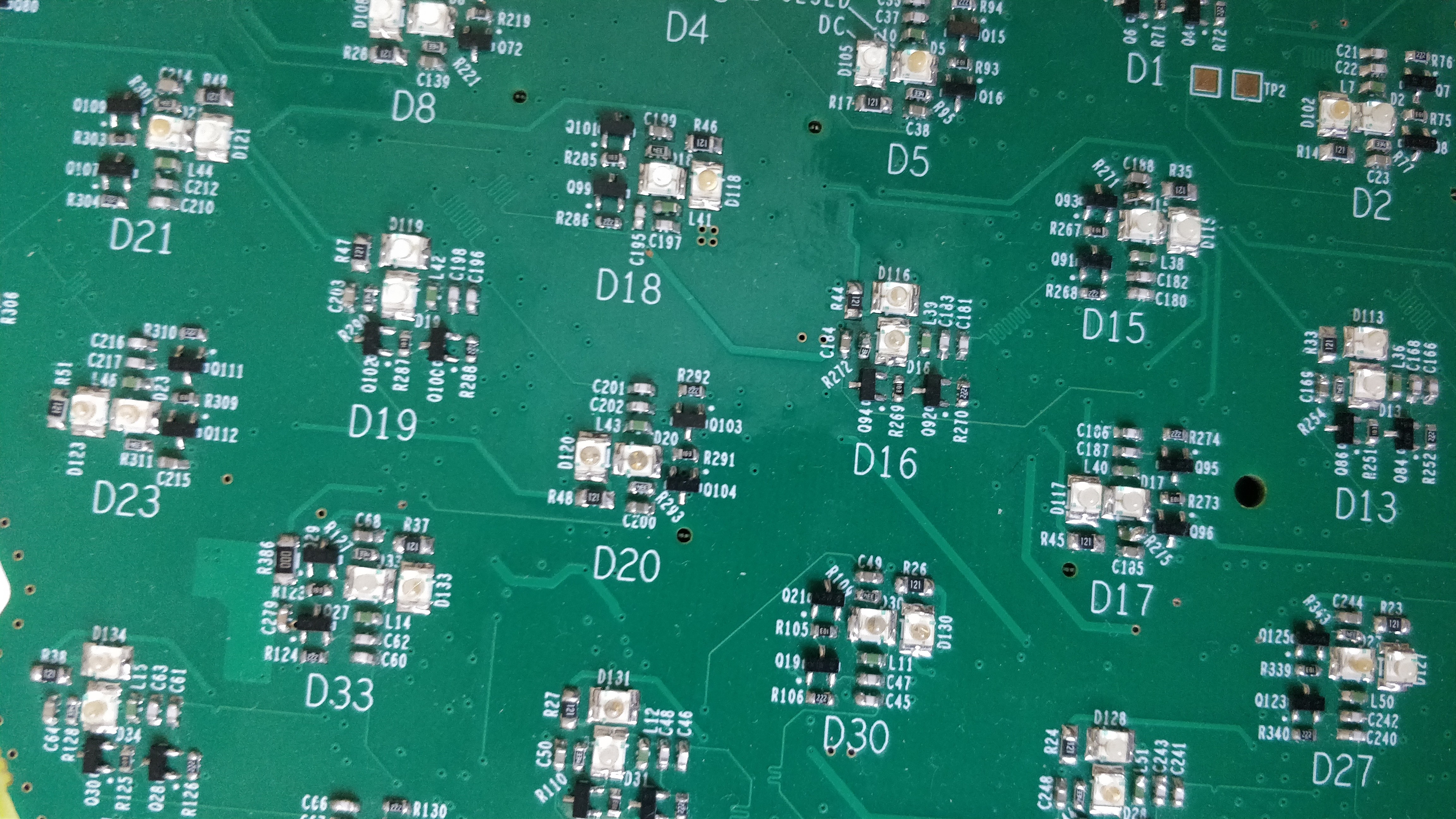}
\caption{From left to right: CTS seen from the LED side, from the back and details of the LED placement\label{fig:CTS}}
\end{figure}

To assess the camera performance and extract all the relevant parameters of its calibration, the Camera Test Setup (CTS), shown in figure~\ref{fig:CTS}, has been developed. The CTS is a hexagonal structure that can be mounted in front of the PDP. It has been conceived to have each pixel of the camera facing two nearby LEDs, one operated in pulsed mode (AC LED), while the other in continuous mode (DC LED). For each pixel, the AC LED is meant to emulate the prompt Cherenkov signal produced by atmospheric showers, while the continuous DC LED emulates the NSB. 
The intensity of the AC LED can be controlled for patches of three neighboring pixels, while the continuous DC LED intensity is adjustable by groups of 24 pixels. 
The light intensity of the LED is controlled by a 10~bit DAC. 
The AC and DC LEDs have both a peak wavelength of $\lambda = 470$~nm (Broadcom / Avago ASMT-BB20-NS000). 
The AC LED is operated from $0$ up to $\sim 8000$~photons (about $2000$~p.e.), while the DC LED is operated to emulate a NSB with frequency up to 1~GHz per pixels.

The CTS is fully programmable and controllable via an OPC-UA server controlled by scripts written in the ALMA Control Software~\cite{ALMA-control-software} (ACS) framework.
Its programmable feature allows injecting different kinds of patterns; therefore also simulated images of showers or muon rings can be injected in the CTS to test the camera performance and trigger algorithms. 

\begin{figure}[htbp]
\centering
\includegraphics[width=0.9\textwidth]{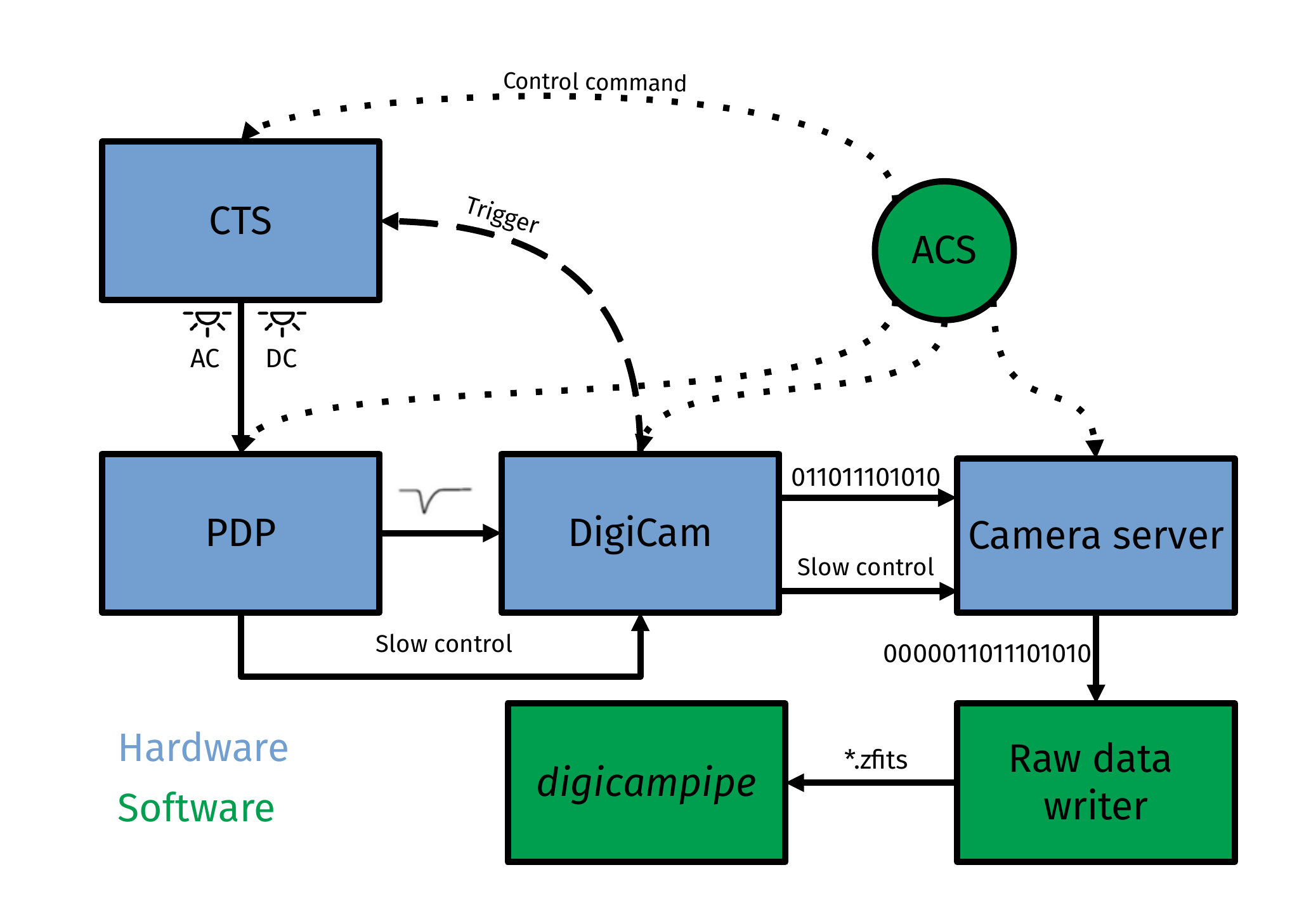}
\caption{Schematic of the CTS data acquisition setup. \label{fig:CTS_setup}}
\end{figure}

The trigger signal and the communication bridge are provided to the CTS by DigiCam. Its mechanics have been optimized such that the CTS can be mounted on the camera without removing the shutter doors nor the protection window and still providing light tightness. 
The light tightness of the CTS allows to re-calibrate the camera on-site without needing to remove it from the telescope structure, thus saving on the maintenance time.

\subsubsection{Acquisition and processing}\label{sec:acquisition}

The data are acquired as if the camera was mounted onto the telescope during standard observations: The camera is connected via the two 10~Gbps optical fibers (one for the data, one for the slow-control) to the camera server, where the acquisition software runs. All data acquisition components are controlled via ACS by the use of ACS scripts. This feature allows programming all steps of the measurement, from the light generation to the data recording. All measurements performed with the CTS use DigiCam as a trigger source that offers stable synchronization (estimated to be $\leq 100$~ps) between the light pulses and the data acquisition. In addition to the event number, timestamp, and 1296 waveforms, each event contains:

\begin{enumerate}
    \item 16~bit unsigned integers providing the baseline for each pixel computed as the average of the 1024 samples preceding the trigger decision (see equation \ref{eq:baseline_comp});
    \item the sum of the baseline-subtracted waveforms from three neighboring pixels clipped at 8~bit;
    \item 1~bit waveforms per cluster (a group of 7 triplets or 19) and per trigger algorithm, identifying when in the readout window the trigger condition was satisfied.
\end{enumerate}
 
The first item provides a precise measurement of the baseline while the second and third allow studying the trigger. For example, they allow to measure the trigger efficiency as both external and internal triggers can be flagged as true for the same event.

The data analysis is performed using a dedicated software package \textit{digicampipe}~\cite{digicampipe} developed by the SST-1M team.\footnote{It has been constructed to be based on the core components of \textit{ctapipe}~\cite{CTA-ctapipe-ICRC19}, the reconstruction pipeline software for CTA.} The code contains calibration scripts, reconstruction and data quality pipelines for the SST-1M telescope. A schematic of this acquisition setup is shown in figure~\ref{fig:CTS_setup}.

\paragraph{AC/DC scan}\label{sec:ACDC_scan}

\begin{figure}[htbp] 
\centering 
\includegraphics[width=0.45\textwidth]{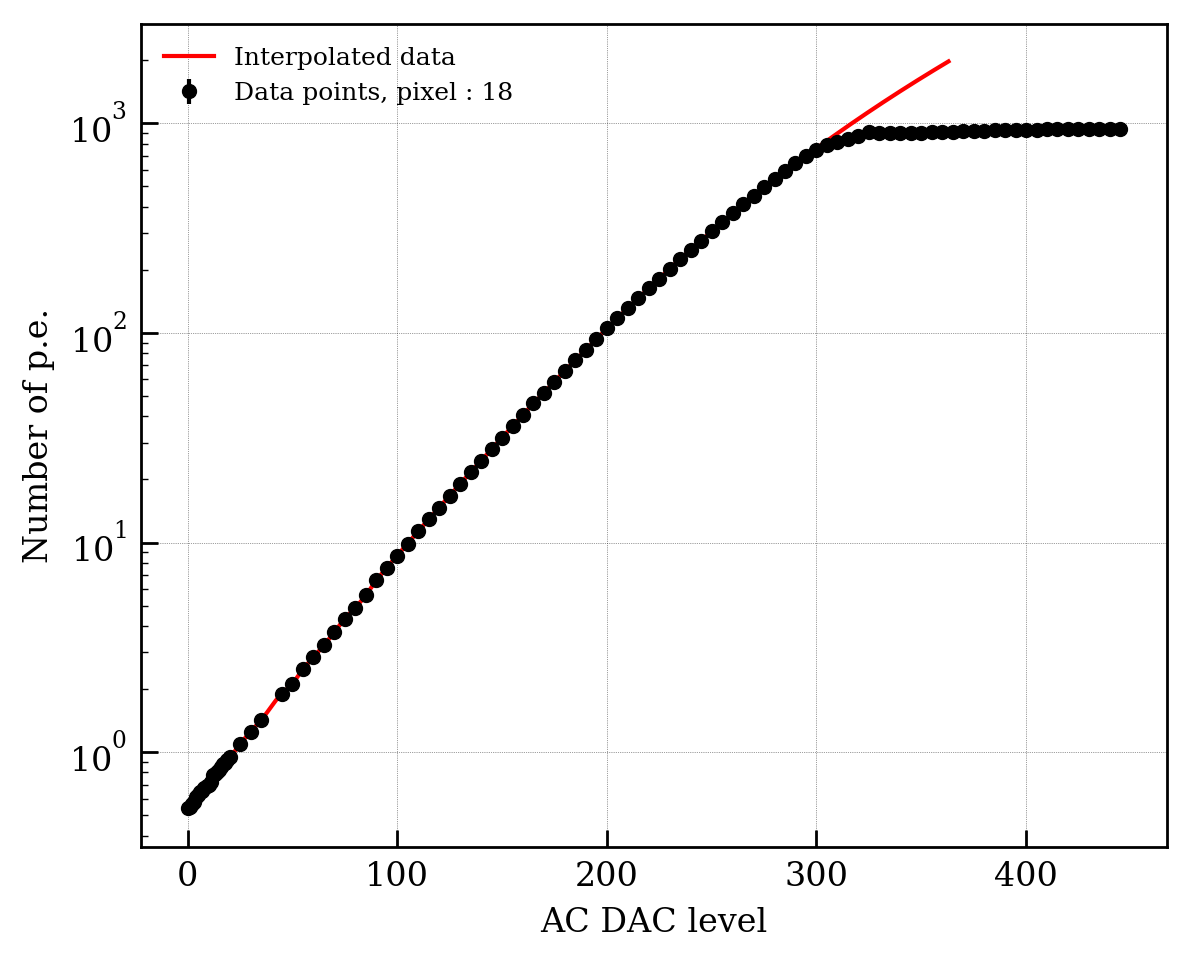}\hfill \includegraphics[width=0.45\textwidth]{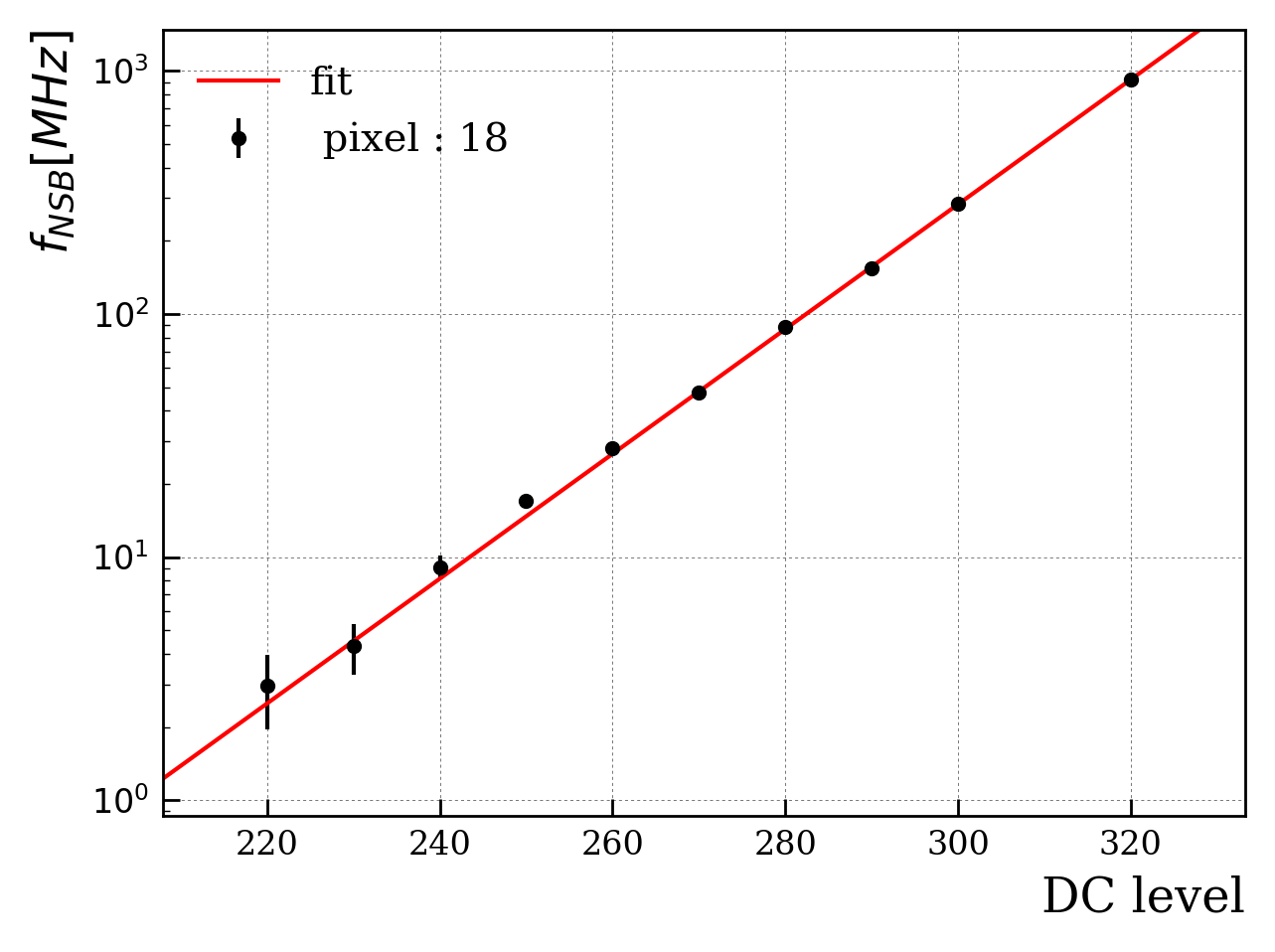}\hfill \caption{Left: Pulsed CTS LED (AC LED) calibration curve for the LED pair numbered 18. The number of photo-electrons is obtained from the pulse amplitude in the linear regime of the amplification chain. They were extrapolated using a forth degree polynomial above saturation.  Right: Continuous CTS LED (DC LED) calibration curve for the LED pair numbered 18.}\label{fig:leds}
\end{figure}

A scan of pulsed and continuous light-level is realized to infer the photo-sensor properties. A set (referred to as AC/DC scan in the following) of $N_{\textrm{w}}=10^4$ waveforms, per pixel, for each level of pulsed and continuous light (AC and DC light) is acquired with the CTS. The AC light ranges from 0 up to $\sim 2 \times 10^4$ photons in $N_{\textrm{AC}}=50$ levels, while the DC light ranges from 0 up to $\sim 1$~GHz equivalent night-sky background rate in $N_{\textrm{DC}}=10$ levels. The scan without continuous light is used to determine the reference SiPM characteristic parameters. 
Using the single photon-counting capabilities of SiPMs, the average number of photo-electrons $\mu_j$ for the $j^{\rm th}$ pulsed-light level can be determined allowing to calibrate the AC LEDs at the same time. 
The LEDs have been calibrated using the SiPM sensors in the pre-amplifier linear range ($N_{\textrm{p.e.}} \leq 600$~p.e.).
In the non-linear range, the calibration of the LEDs is extrapolated using a calibrated photo-diode as described in~\cite{SST-1M-camera}. Figure~\ref{fig:leds} shows the calibration of a pair of LEDs. 

\paragraph{Time delay scan}\label{sec:timing_scan}

Compared to the sampling frequency $f=250$~MHz of the FADCs, the full width at half-maximum of the photo-electron pulse of $\sim 20$~ns is quite short, meaning that the rising and falling edges of the pulse typically fall within 5 samples (each of 4~ns). 
Therefore, for a given time delay between the flash of light and the readout window, the pulse waveform might not be sampled at its maximum. 
To artificially obtain a higher sampling frequency, the delay between the trigger and the readout window can be adjusted in DigiCam by steps of 77~ps over 32 steps.
This allows to shift up to 2.464~ns (32$\times$77~ps) the readout and trigger windows within the 4~ns sample of the FADC, therefore capturing nearly all features of the waveform. 
Low jitter and stable light source are required to be able to combine the waveforms for the different delays and obtain a precise pulse template for each pixel. This scan is referred to as the time delay scan in the following. 

\subsubsection{Analysis and results}

\paragraph{Single photo-electron spectrum}\label{sec:spe}

\begin{figure}[htbp]
    \centering
    \includegraphics[width=0.7\textwidth]{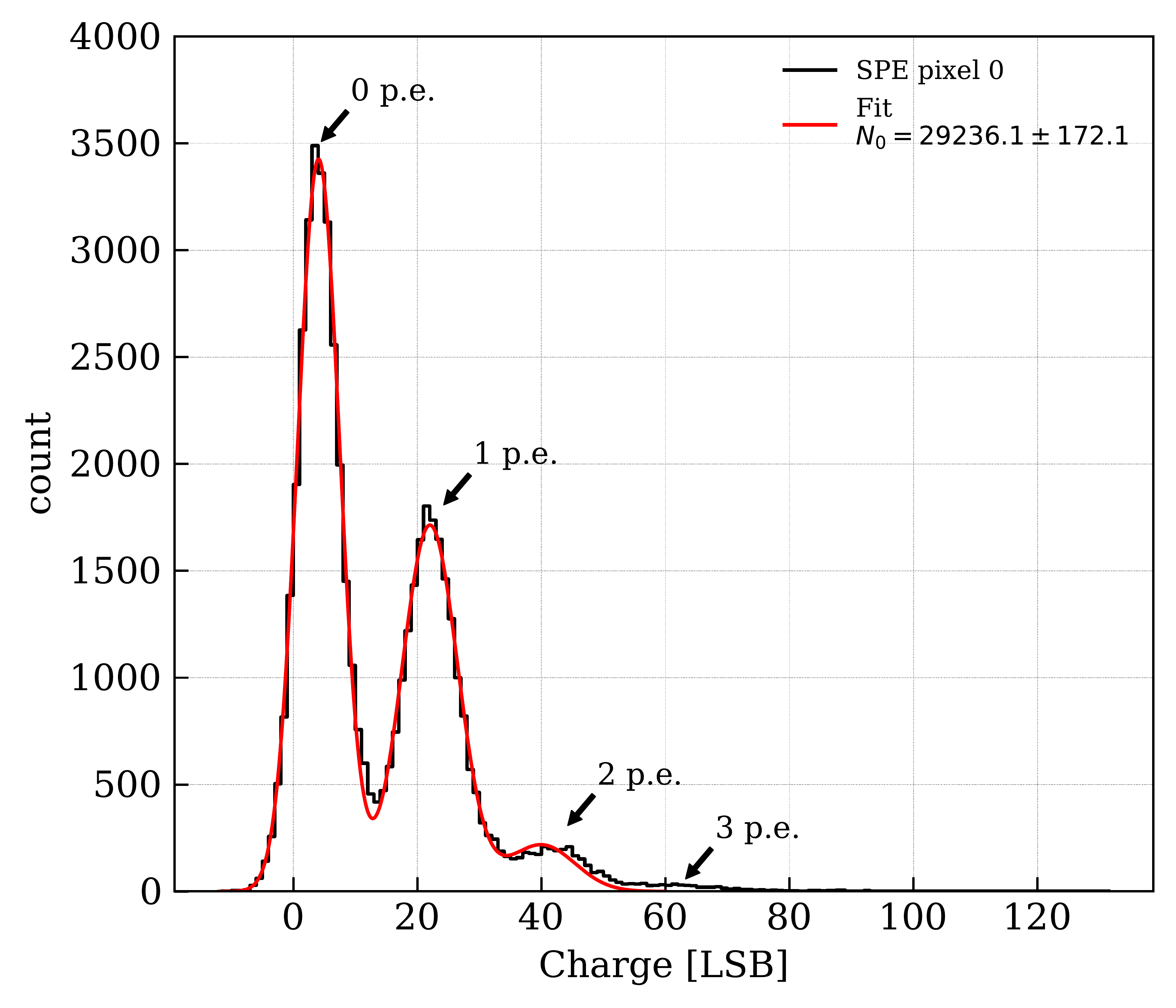}
    \caption{Example of a single photo-electron spectrum fitted with equally spaced Gaussians.}
    \label{fig:spe_exemple}
\end{figure}

The single photo-electron spectrum (SPE) is derived from the charge distribution within the readout window in absence of external light source, therefore containing only thermal pulses. It is obtained in dark conditions, namely in the absence of AC and DC light. An example of a SPE is shown in figure~\ref{fig:spe_exemple}.

\begin{figure}[htbp]
    \centering
    \includegraphics[width=0.45\textwidth]{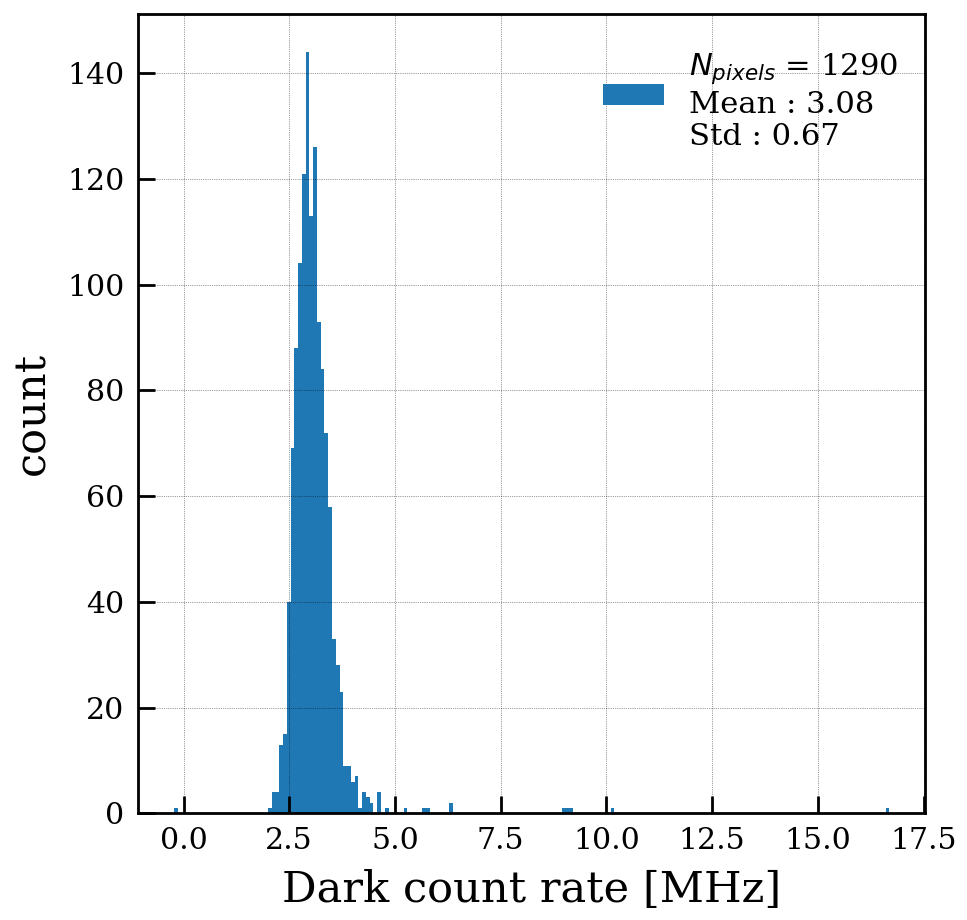}
    \includegraphics[width=0.45\textwidth]{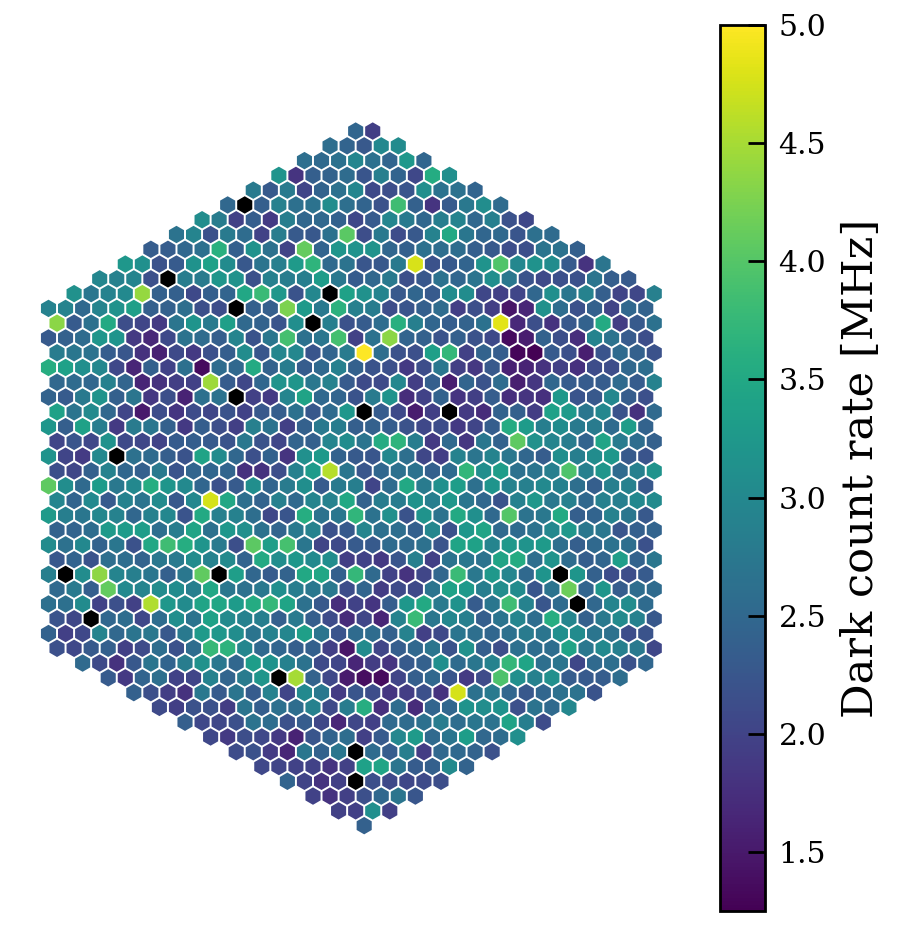}
    \caption{Left: Dark count rate distribution over the camera pixels. Right: Dark count rate viewed for each pixel.}
    \label{fig:spe}
\end{figure}

From the SPE, the number of events $N_0$, corresponding to the number of times no dark pulses were observed within the readout window, is obtained by fitting the SPE with equally spaced Gaussians. Using Poisson statistics the dark count rate (DCR) is given by (as in~\cite{SST-1M-SiPM}):

\begin{equation}
    DCR = \frac{<N_{\textrm{pulses}}>}{T_{\textrm{window}}} = \frac{-\ln\left( \frac{N_{\textrm{w}}}{N_0} \right)}{\frac{N_{\textrm{samples}}}{f}},
\end{equation}

where $<N_{\textrm{pulses}}>$ is the average number of dark pulses in a duration $T_{\textrm{window}}$, $N_{\textrm{w}}$ is the number of waveforms acquired and $N_{\textrm{samples}}$ is the number of time samples recorded in each waveform.
An average dark count rate of $3.08$~MHz per pixel is observed (see figure~\ref{fig:spe} (left)).\footnote{Corresponding to $32.9~{\rm kHz/mm^2}$ at $26.5~{\rm ^{\circ}C}$} On the right of this figure, the dark count map over the plane of the camera is shown. A lower dark count rate can be noticed at the lower left edge of the camera, corresponding to a different batch production of the SiPMs. As expected, the dark count rate is correlated with temperature, as can be seen from figure~\ref{fig:glob_fit_res_map} (bottom right) and figure~\ref{fig:spe} (right). 

It should be noted that the dark count rate, as defined above, actually measures the number of thermal pulses above 0.5~p.e. Thus this value also includes after-pulses induced by dark counts with amplitude higher than 0.5~p.e. As the after-pulse probability is of the order of 2\%~\cite{SST-1M-SiPM} and the readout window used is long enough ($T_{\textrm{window}} = 364$~ns) the dark count rate is only slightly overestimated as can be seen from figure 15 in~\cite{SST-1M-SiPM}. Since the dark count rate is about 13 times lower than the expected NSB rate at a high-quality observational site (e.g. the NSB rate in a moon-less night is about 40~MHz in Paranal, Chile), it is not relevant to obtain an DCR measurement unbiased by afterpulses. Moreover, the results are compatible with the ones reported in~\cite{SST-1M-SiPM}. 

The rather high dark count rate observed in the SiPM, due to its large active surface of $93.6$~mm$^2$, makes it difficult to extract with high precision the optical cross-talk and the gain. This is due to the high probability of having coincident uncorrelated dark count events. Therefore, to extract the calibration parameters the multiple photo-electron spectrum (MPE) is used as presented in the following section.

\paragraph{Multiple photo-electron spectrum}\label{sec:mpe}

To extract the gain $G$, the optical cross-talk $\mu_{\textrm{XT}}$, the SiPM gain smearing $\sigma_s$ and the mean number of photo-electrons $\mu_j$ of the $j^{\rm th}$ pulsed light level for each pixel, the single photon counting capability of the SiPM is used. The reconstructed charge $C$ distribution of prompt signals from the AC LED can be expressed with a smeared generalized Poisson distribution as in~\cite{SiPM-Vinogradov}:

\begin{eqnarray}\label{eq:mpe}
    P(C_j=x) &=&  \sum_{k=0}^{\infty} \frac{\mu_j (\mu_j + k\mu_{\textrm{XT}})^{k-1}}{k!} e^{-\mu_j -k\mu_{\textrm{XT}}}\frac{1}{\sqrt{2\pi}\sigma_k} e^{-\frac{(x - kG - \bar{B})^2}{2\sigma_k^2}} \\
    & {\rm with} & G = \bar{G}\int_0^{\Delta t} h(t)dt, \label{eq:gain_conversion} \\
    & & \sigma_k^2 = f\Delta t\sigma_e^2 + kG\sigma_s^2, 
\end{eqnarray}

\noindent
where $\sigma_e$ is the electronic noise per time slice, $\bar{G}$ is the amplitude of a single photo-electron signal and $\bar{B}$ is a free parameter allowing to center the first photo-electron peak to a null value, that can be interpreted as the residual charge after baseline subtraction as defined in section~\ref{sec:photon-reco}.

A maximum log-likelihood estimation (MLE) on a smeared generalized Poisson distribution is applied to the charge spectrum (or multiple photo-electron spectrum) of each pulsed light level $j$ acquired, where the log-likelihood is defined as:

\begin{equation}\label{eq:llh_reduced}
   \ln\mathcal{L}(\vec{\theta}) = \frac{1}{N_{\textrm{w}}} \sum_{i=1}^{N_{\textrm{w}}} \ln P (\vec{\theta}; C_{ij}),
\end{equation}

with $P(\vec{\theta}; C_{ij})$ defined from equation~\ref{eq:mpe} and represents the likelihood of the parameters $\vec{\theta}$ given the observed charge $C_{ij}$ of the $i^{\rm th}$ waveform and the $j^{\rm th}$ AC light level.
All fit parameters $\vec{\theta}$, except the mean number of photo-electrons $\mu_j$, are not changing with the amount of light emitted by the LEDs. Therefore the fit results for each input pulsed light level should be consistent. To achieve that the log-likelihood $\ln \mathcal{L}$ is further reduced by summing over the pulsed light level $j$:

\begin{equation}\label{eq:llh}
   \ln \hat{\mathcal{L}}(\vec{\theta}) = \frac{1}{N_{\textrm{w}}N_{\textrm{AC}}} \sum_{j=1}^{N_{\textrm{AC}}} \sum_{i=1}^{N_{\textrm{w}}} \ln P(\vec{\theta}; C_{ij}).
\end{equation}

By combining the MLE of each light level, the number of fit procedures is reduced from $N_{\textrm{AC}}$ to $1$ and the number of independent fit parameters from $6 \times N_{\textrm{AC}}$ to $N_{\textrm{AC}} + 5$. The number of fit parameters can be again drastically reduced by imposing that the sample mean $<C_{ij}>_i$ and the mean number of photo-electrons $\mu_j$ follow the relation derived from the mean of the smeared generalized Poisson distribution (in equation~\ref{eq:mpe}):

\begin{equation}
    \mu_{j} = \frac{<C_{ij}>_i - \bar{B}}{G} (1 - \mu_{\textrm{XT}})
\end{equation}

The overall reduction and simplification of the MLE increases the robustness of the fit and reduces the chances of falling into local minima. The number of independent fit parameters is, in the end, reduced to $5$ (the gain $G$, the optical cross-talk $\mu_{\textrm{XT}}$, the electronic noise $\sigma_e$, the gain smearing $\sigma_s$ and the residual baseline $\bar{B}$). It allows to fit consistently the MPE from the lowest light levels, where the photo-electron peaks are clearly distinguishable, up to the highest light level, where one cannot resolve single photo-electrons. 

\begin{figure}[htbp]
    \centering
    \includegraphics[width=0.45\textwidth]{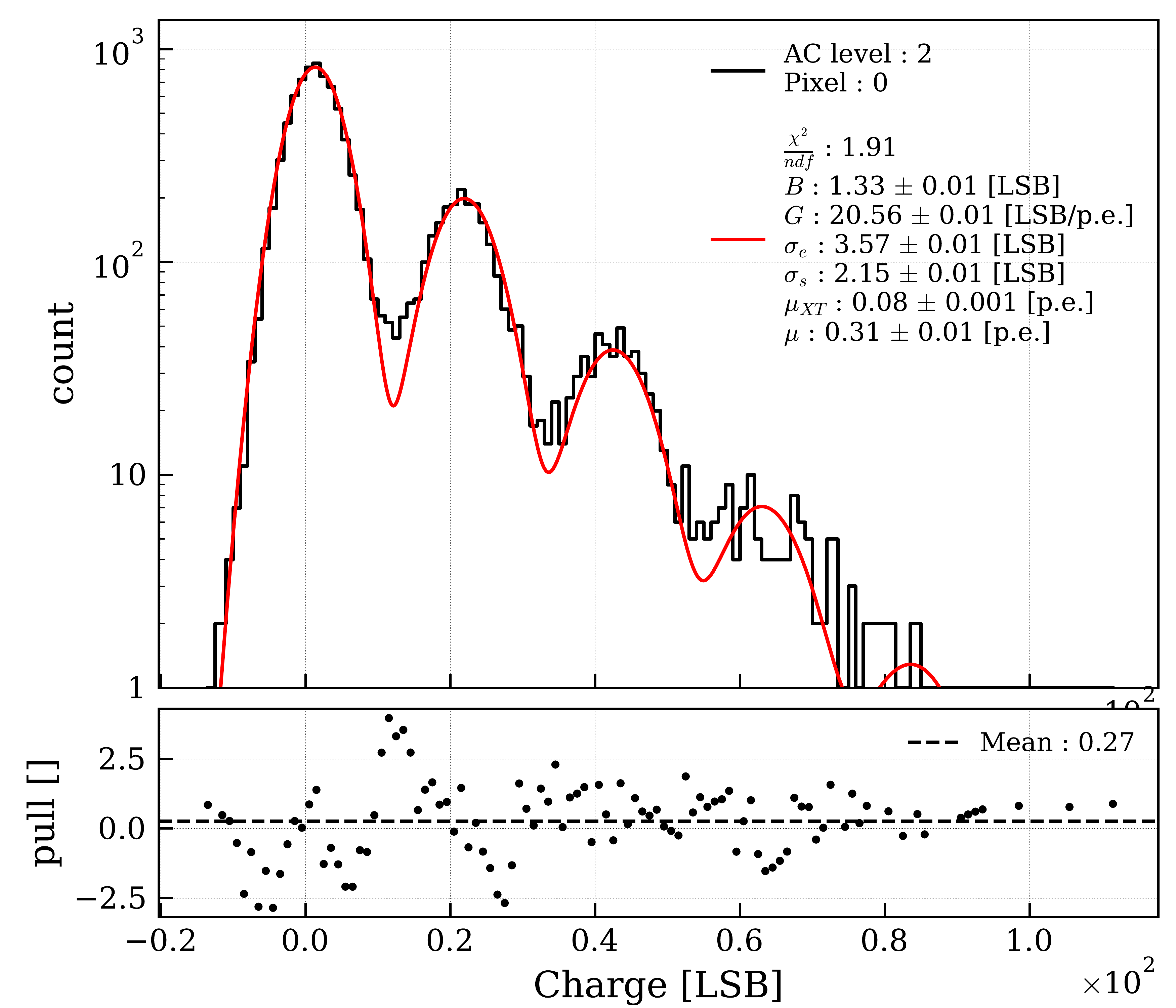}
    \includegraphics[width=0.45\textwidth]{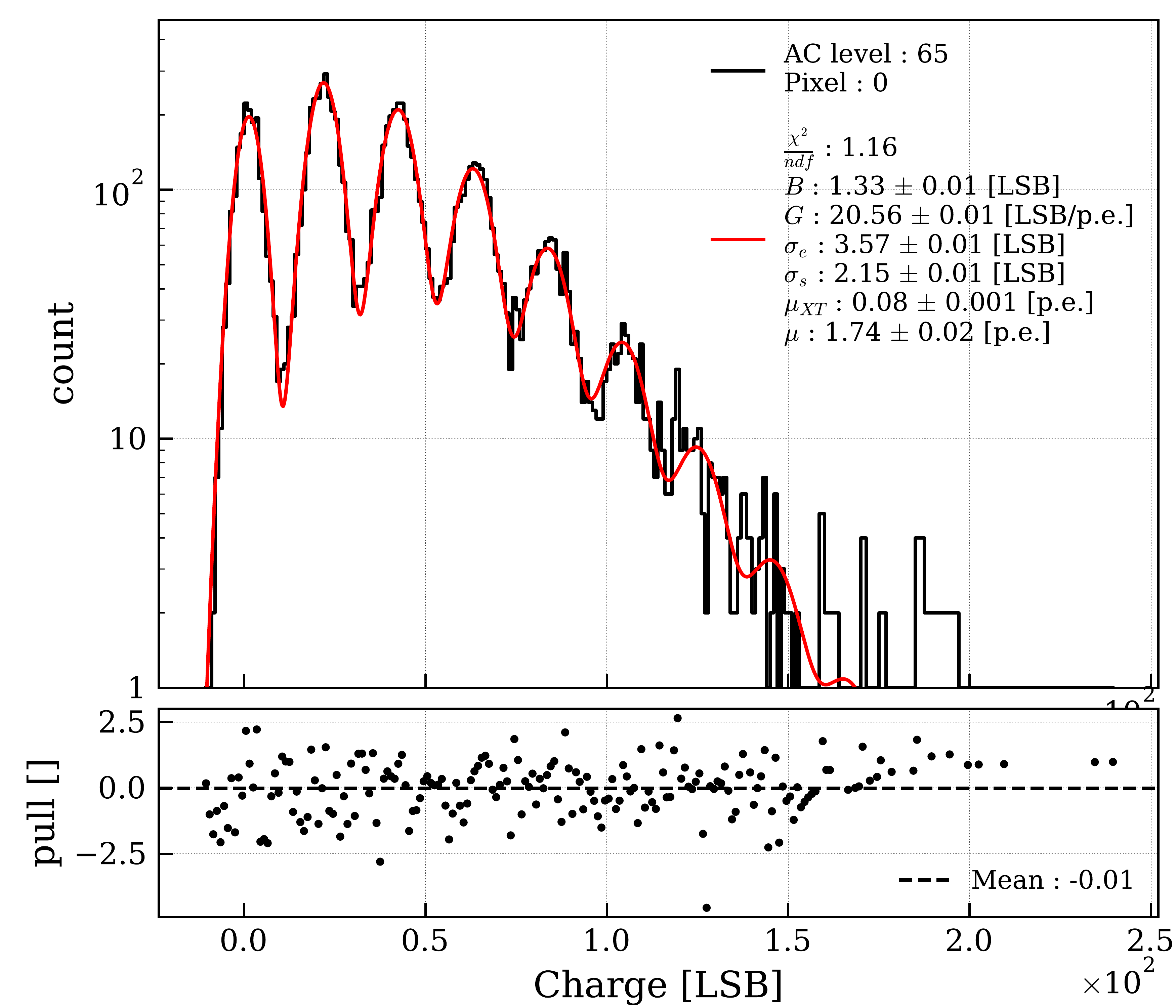}
    \includegraphics[width=0.45\textwidth]{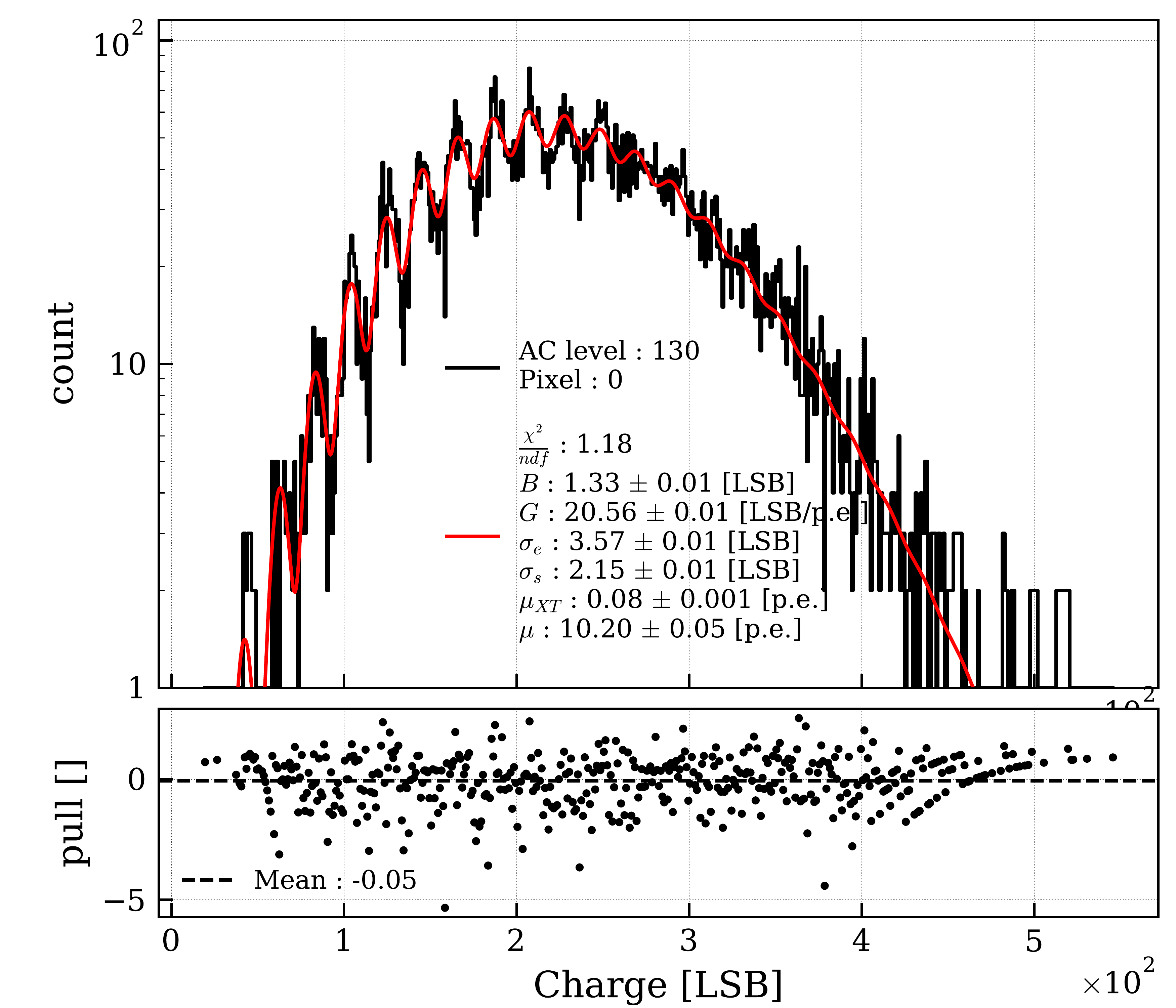}
    \includegraphics[width=0.45\textwidth]{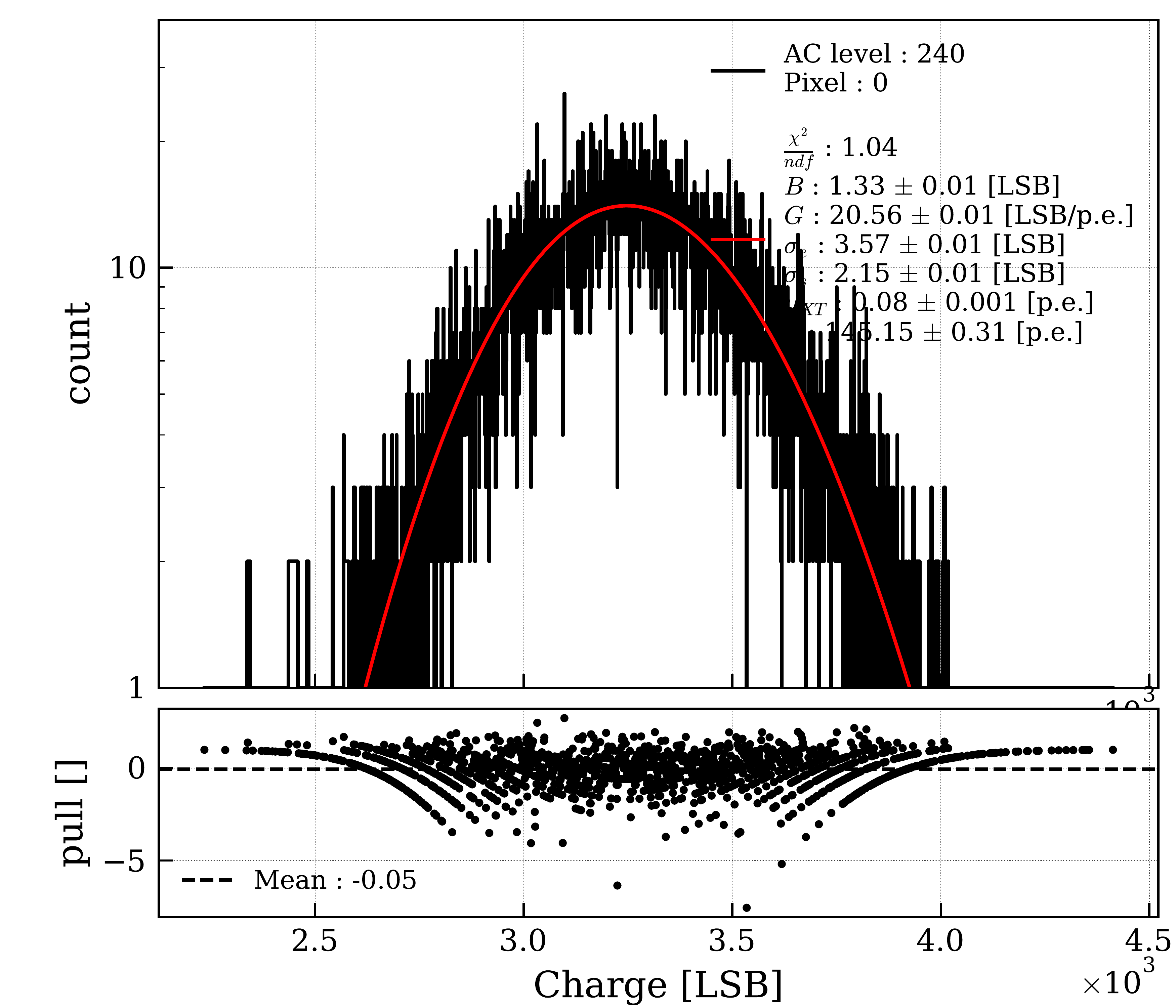}
    \caption{Multiple photo-electron distributions, of pixel 0,  acquired with different light levels. It shows four distinct light levels: 0.31~p.e. (top left), 1.74~p.e (top right), 10.2~p.e. (bottom left) and 145.15~p.e. (bottom right) fitted with the probability density function defined in equation~\ref{eq:mpe}. The fit for these light levels is performed in a combined way by maximizing the log-likelihood defined in equation~\ref{eq:llh_reduced}. The reduced chi-squared for each fitted spectra is reported individually as an estimate of the goodness of fit.}
    \label{fig:global_fit}
\end{figure}

Figure~\ref{fig:global_fit} shows four distinct light levels of the same pixel together with the fit results. Only the mean light level $\mu_j$ differs from one distribution to the other, all other parameters being common.

\begin{figure}[htbp]
    \centering
    \includegraphics[width=0.3\textwidth]{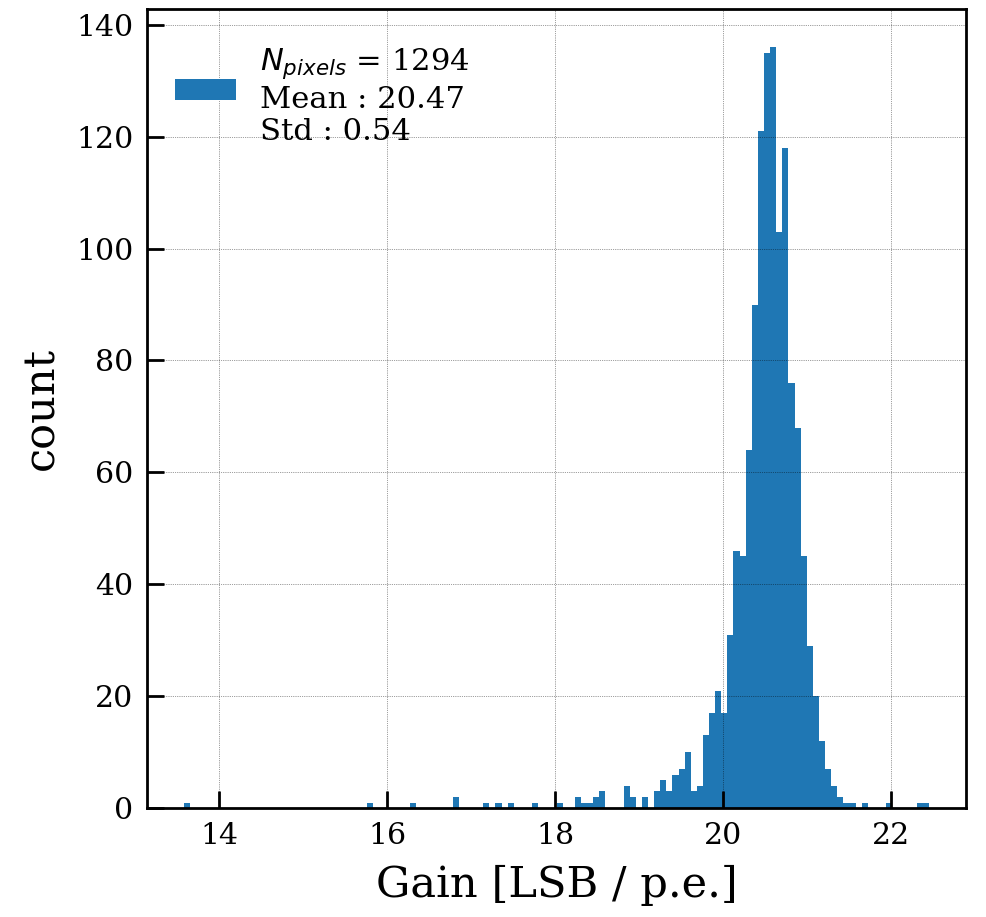}
    \includegraphics[width=0.3\textwidth]{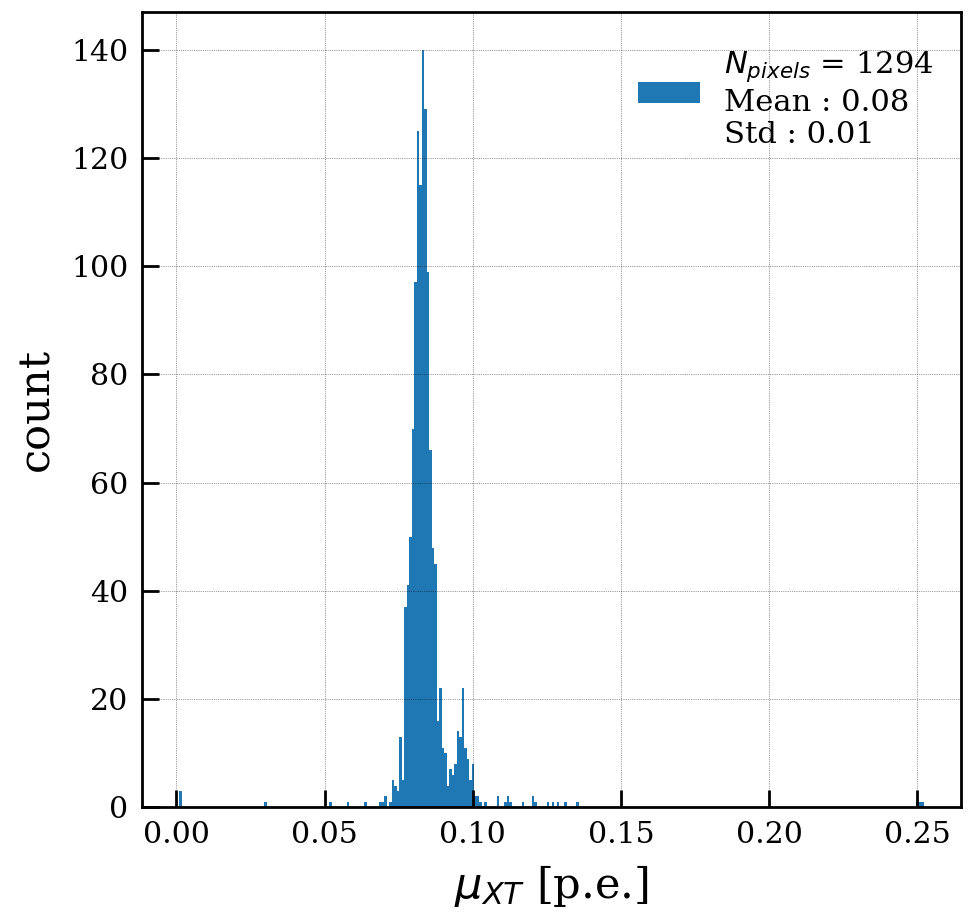}
    \includegraphics[width=0.3\textwidth]{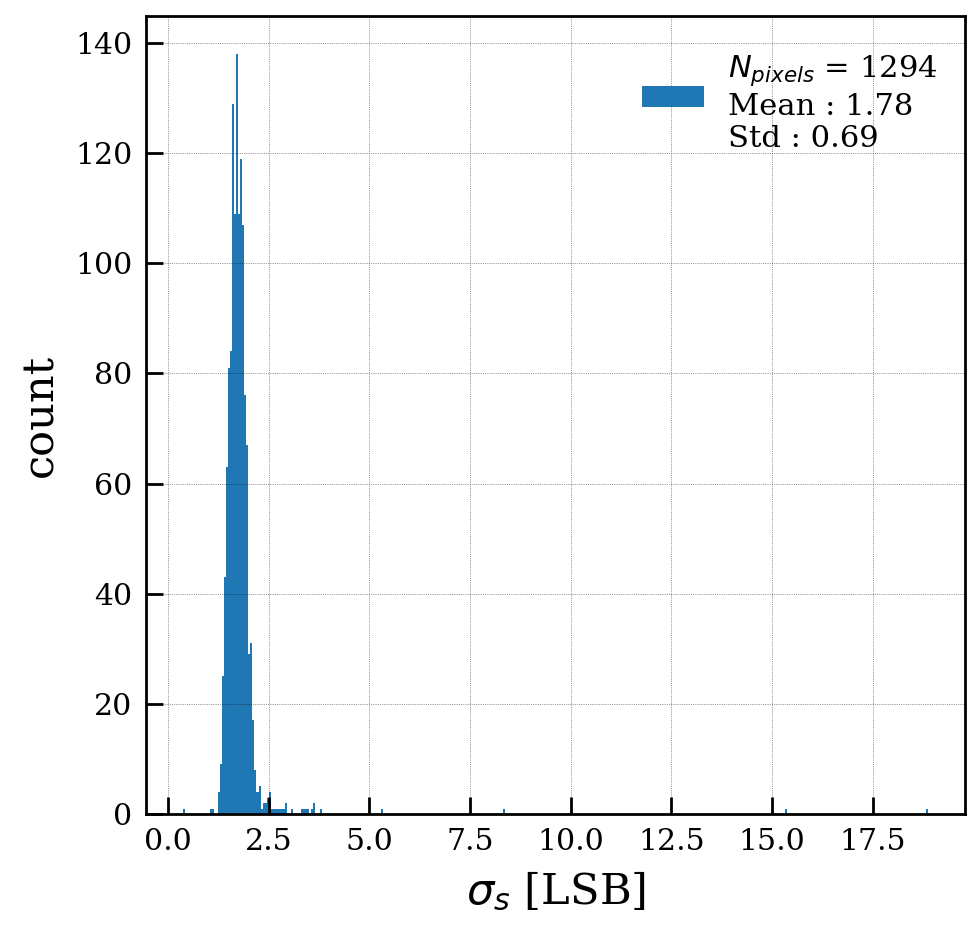}
    \includegraphics[width=0.3\textwidth]{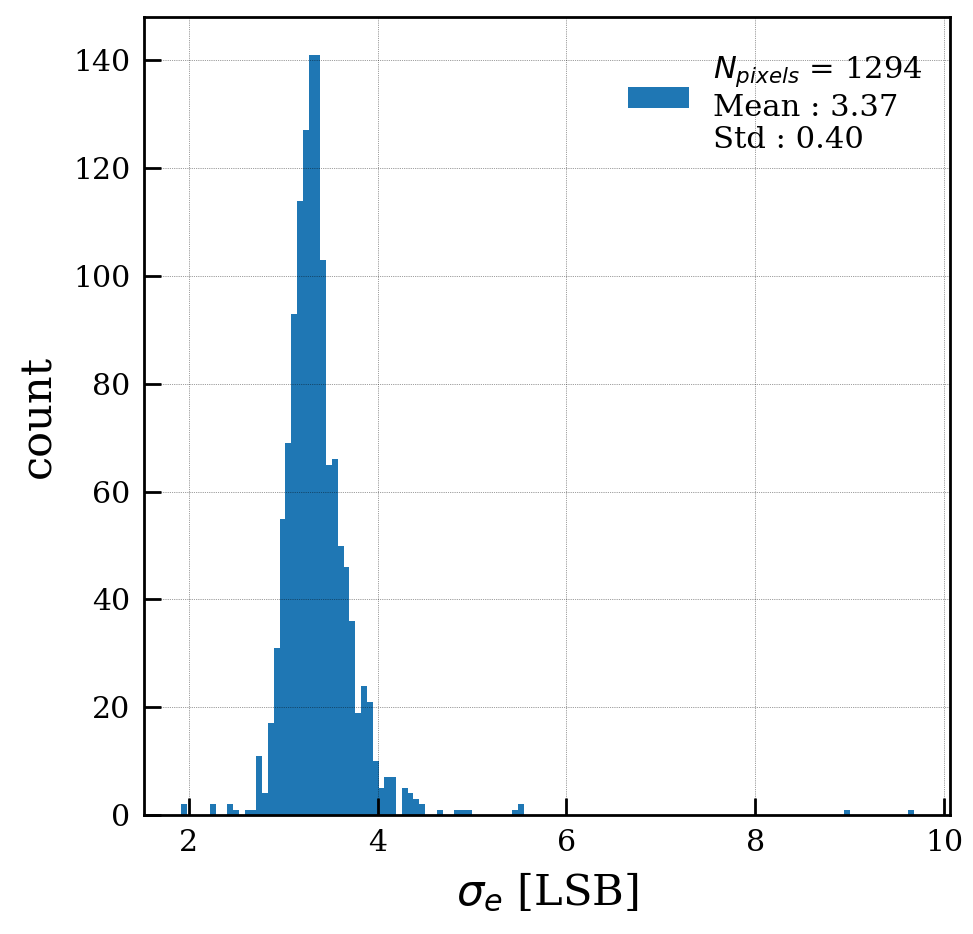}
    \includegraphics[width=0.3\textwidth]{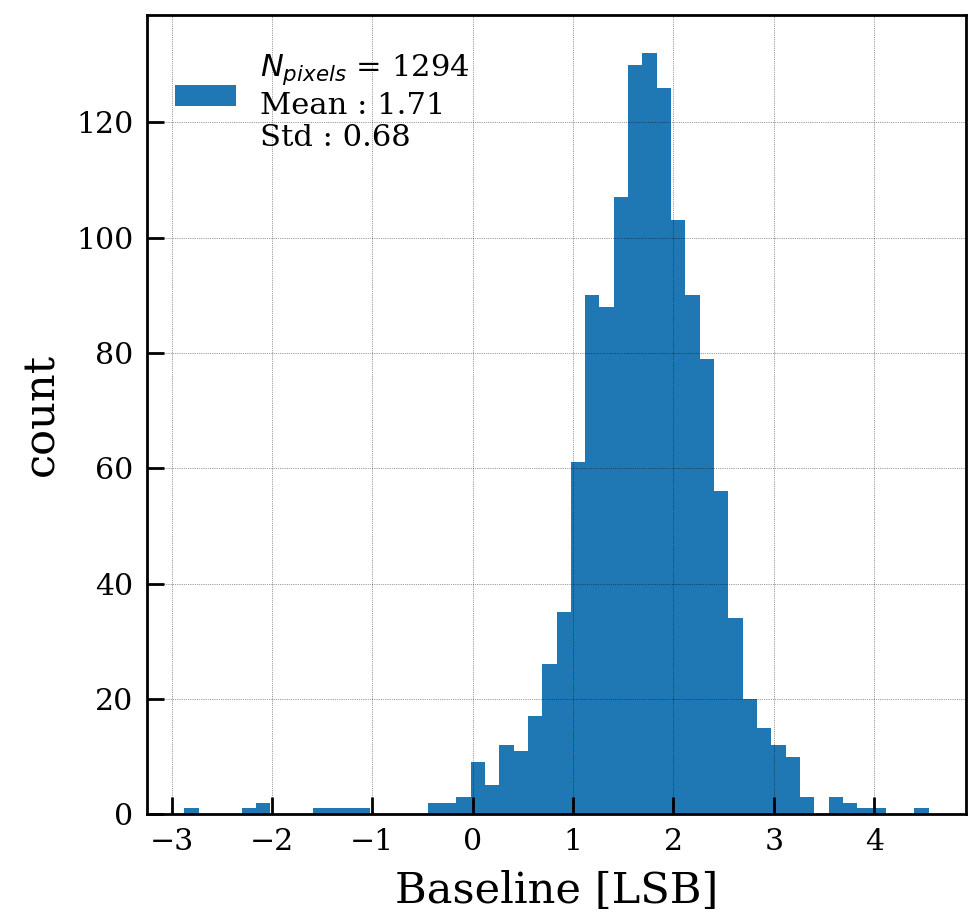}
    \includegraphics[width=0.3\textwidth]{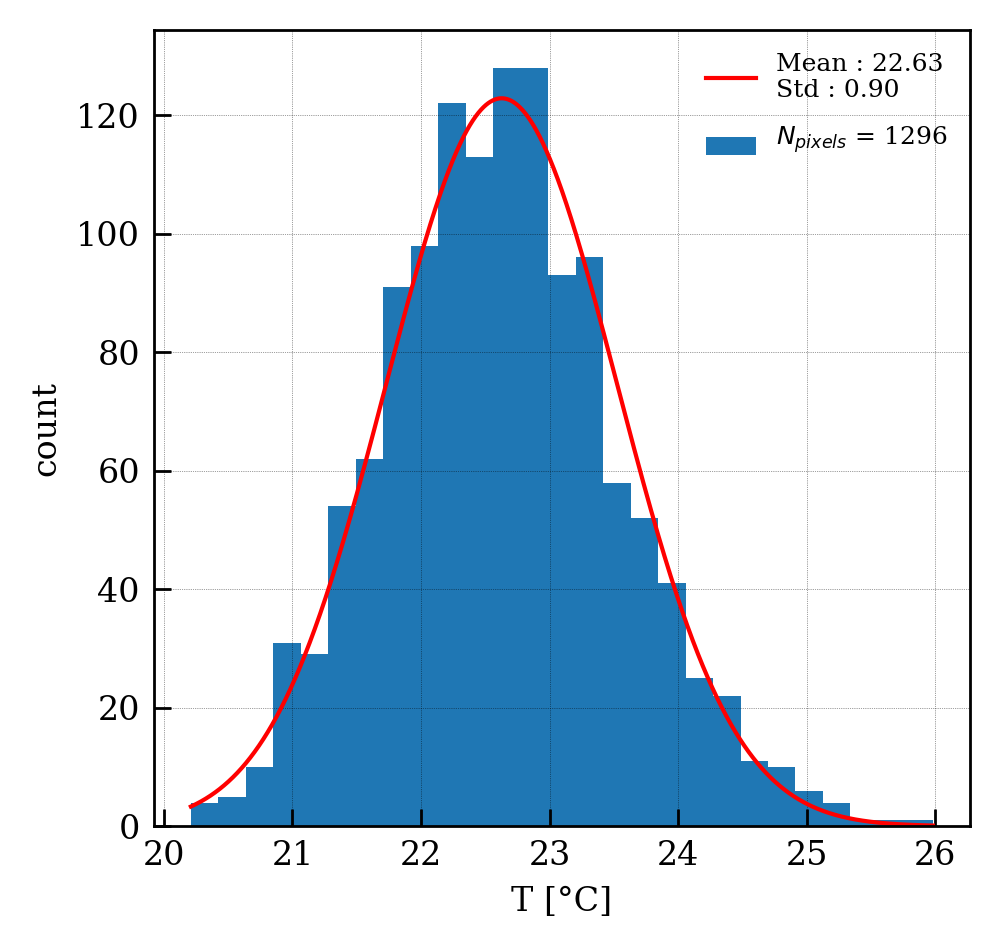}
    \caption{Measured fit parameter distributions for all the 1296 pixels. In reading order, the gain $G$, the optical cross-talk $\mu_{\textrm{XT}}$, the gain smearing $\sigma_s$, the electronic noise $\sigma_e$, the baseline residual $\bar{B}$ and the average temperature of the SiPMs during the measurements are shown.}
    \label{fig:glob_fit_res}
\end{figure}

\begin{figure}[htbp]
    \centering
    \includegraphics[width=0.3\textwidth]{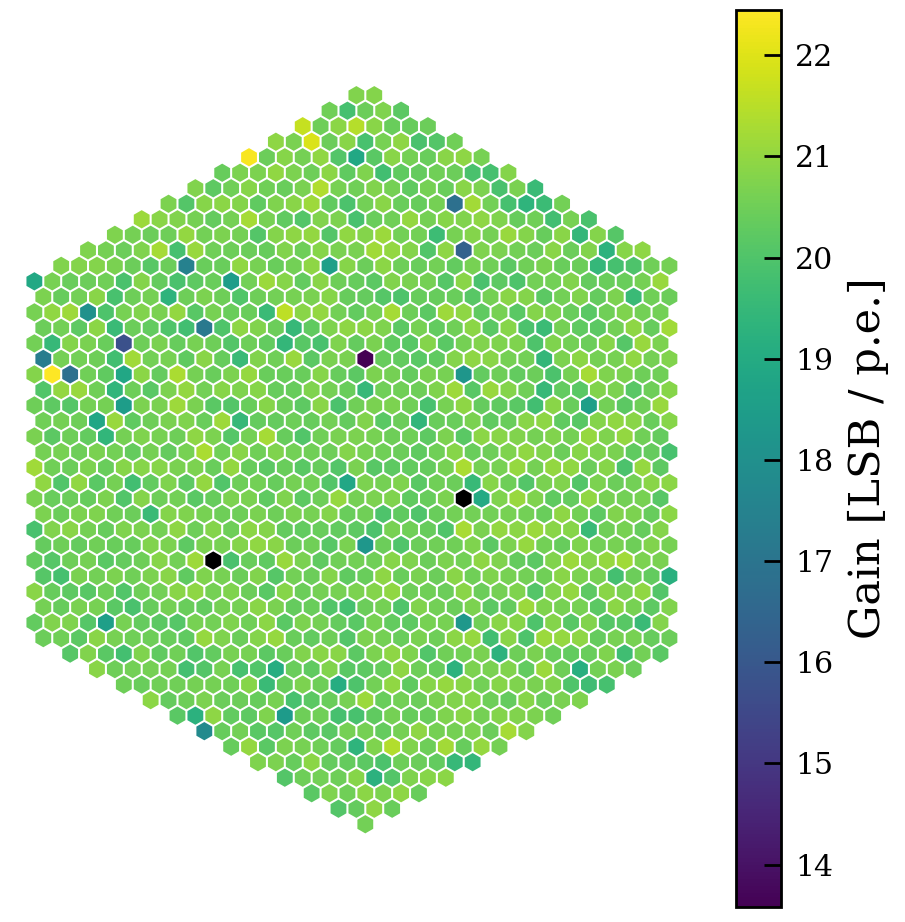}
    \includegraphics[width=0.3\textwidth]{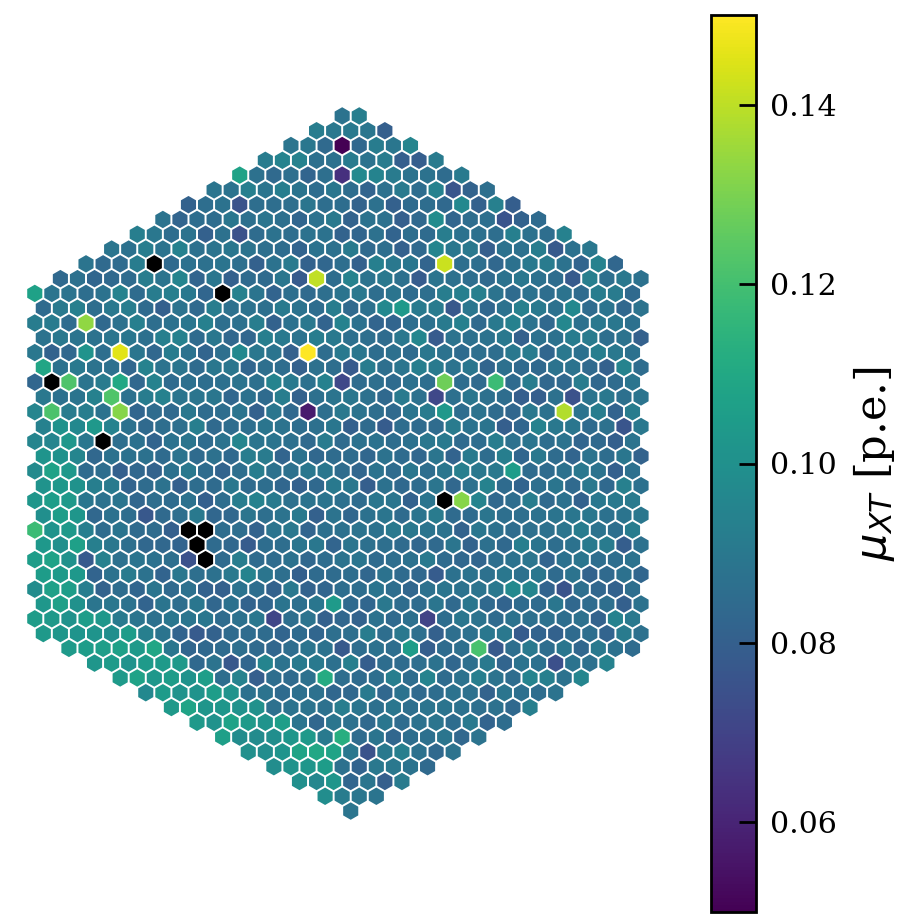}
    \includegraphics[width=0.3\textwidth]{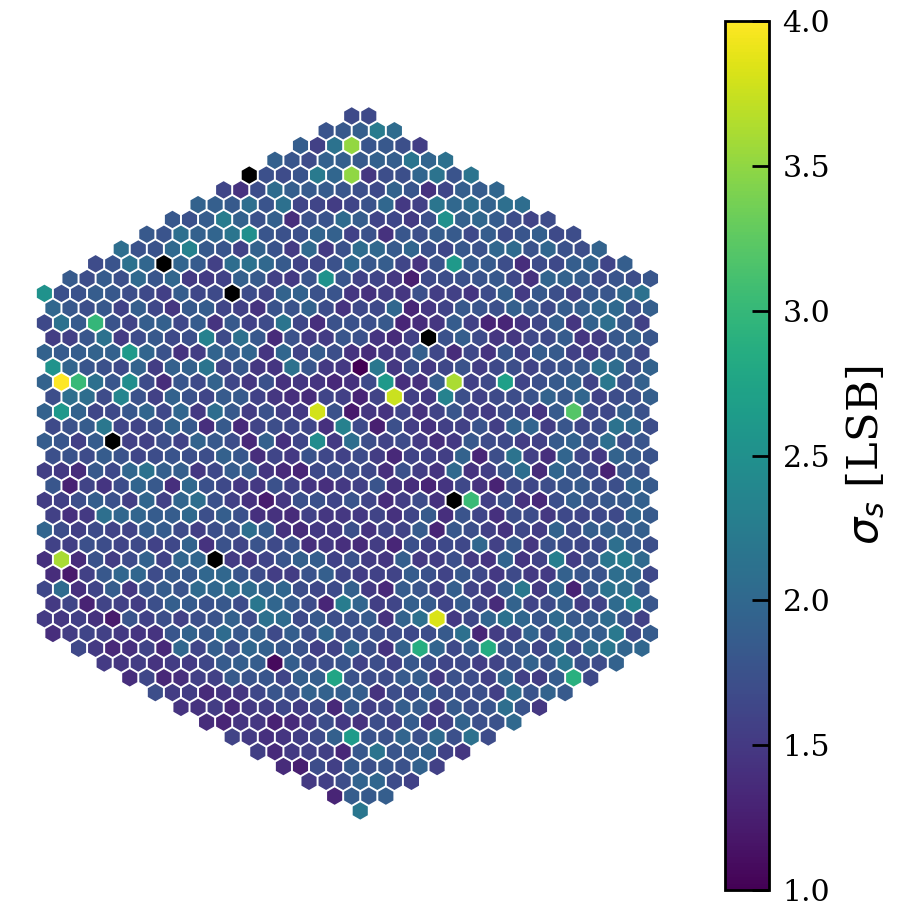}
    \includegraphics[width=0.3\textwidth]{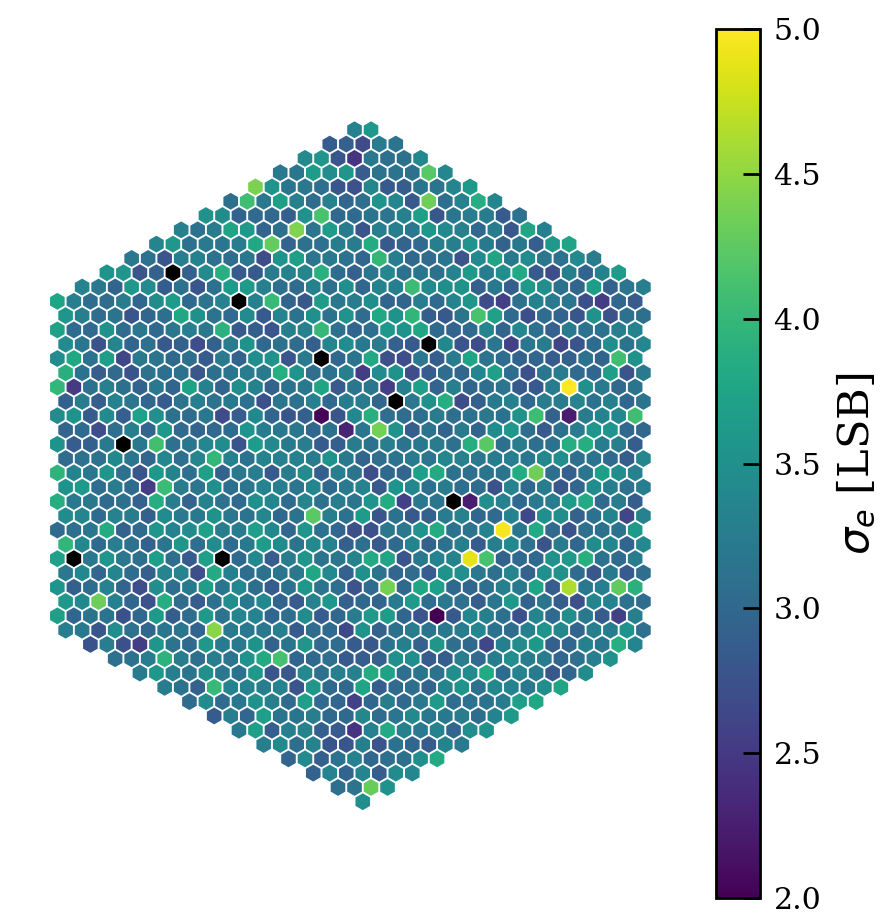}
    \includegraphics[width=0.3\textwidth]{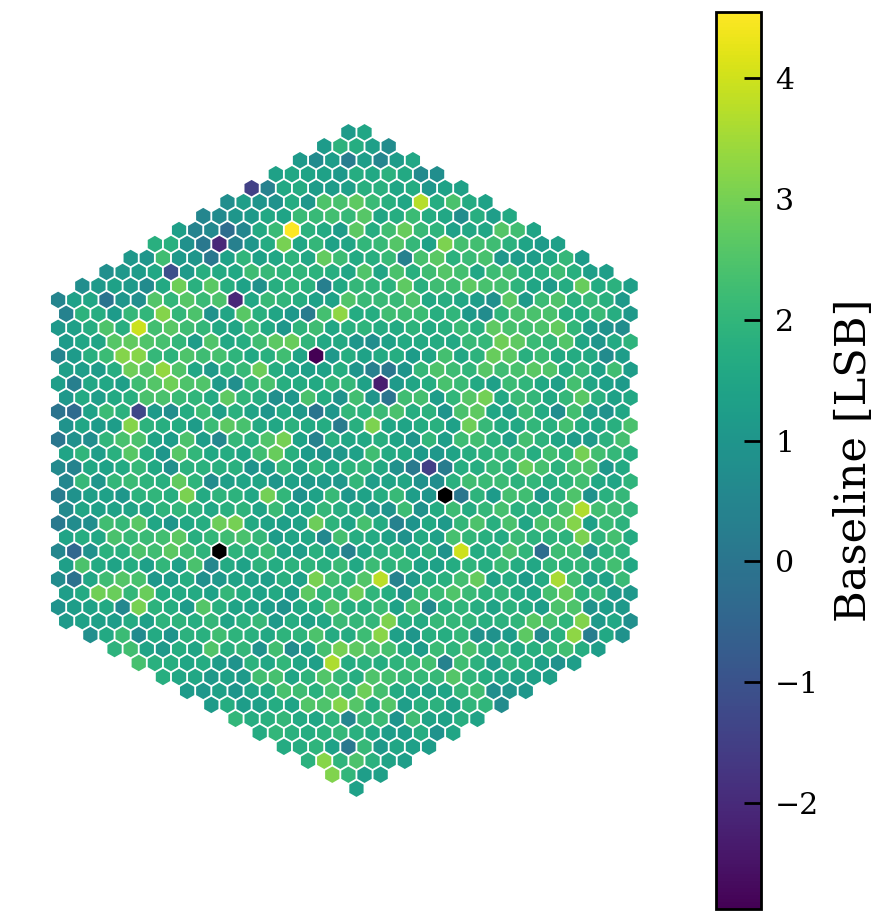}
    \includegraphics[width=0.3\textwidth]{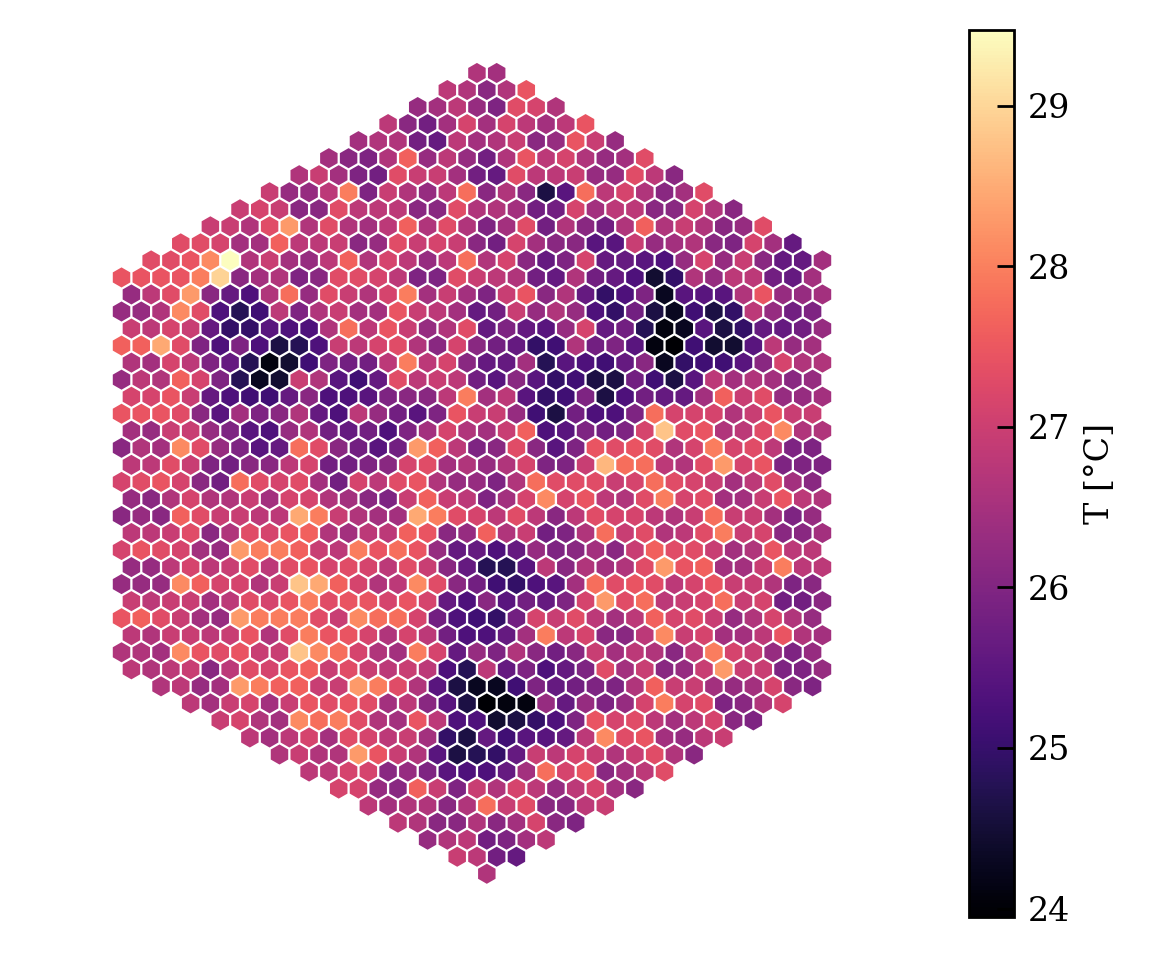}
    \caption{Measured fit parameter maps across the camera. In reading order, the gain $G$, the optical cross-talk $\mu_{\textrm{XT}}$, the gain smearing $\sigma_s$, the electronic noise $\sigma_e$, the baseline residual $\bar{B}$ and the average temperature of the SiPMs during the measurements are shown. The pixels appearing in black are considered as outliers.}
    \label{fig:glob_fit_res_map}
\end{figure}

This method has been applied to all the camera pixels individually. The distributions of the fit parameters are shown in figure~\ref{fig:glob_fit_res} and~\ref{fig:glob_fit_res_map} and summarized in table~\ref{tab:glob_fit_res}. The results were obtained after the first round of gain equalization of the full readout chain using the FADCs variable gain (described in section~\ref{sec:gain_eq}). We observe an average optical cross-talk of 0.08~p.e. (this value corresponds to the average number of cross-talk photo-electrons induced by a single avalanche) which is comparable to the measurements performed in~\cite{SST-1M-SiPM} (10\% optical cross-talk at the operational voltage). A higher optical cross-talk is observed on the left edge of the camera. This part corresponds to the second batch of production of SiPMs. The differences observed are within the producer specifications (typical value 8\% optical cross-talk, maximum value 15\% optical cross-talk). It also highlights the precision at which the optical cross-talk can be determined ($\leq 0.01$~p.e.). 
A similar pattern is observed for the SiPM gain smearing on the same part of the camera. In that case, the gain smearing is lower, consequently the photo-electron resolution ($\sim 0.09$~p.e. on average) is improved. Lower values of gain smearing can be also observed at the camera center because the SiPM with the best photo-electron resolution were mounted on purpose at the camera center for better image resolution. 
An average electronic noise of 16\% of a single photo-electron signal is observed. 
The baseline residual, on average, is of the order of 1.17~LSB, which translates to a systematic underestimation of the baseline evaluated by the DigiCam FPGAs of 5.7\% of the photo-electron amplitude.
This can be explained by the 16~bit limitation of the baseline computation and rounding effects. Nevertheless the systematic baseline offset can be coped off-line by adjusting the baseline value provided by DigiCam accordingly.

The systematic biases of the parameters measured from multiple photo-electron spectra, described in this section, is determined using a Monte Carlo simulation. The Monte Carlo is designed to reproduce the PDP response to a prompt light-source and a continuous background light-source. The Monte Carlo simulates the baseline and its fluctuations due to NSB, dark count rate and electronic noise. It simulates also the amplification chain gain and gain smearing. The Poisson fluctuations and optical cross-talk are also simulated. It is similar to the one described in~\cite{SST-1M-camera} but it includes optical cross-talk as modeled in~\cite{SiPM-Vinogradov}. The result shows that the bias increases with an increased thermal noise rate. In particular, the relative differences between the true and reconstructed parameters for a dark count rate of 3.08~MHz (observed average dark count rate for the off-site calibration) shows negligible systematic error and good resolution in the reconstructed parameters (see table~\ref{tab:systematics}). Moreover, the bias induced in the reconstructed number of photo-electrons, from equation~\ref{eq:pe_conversion}, is of the order of $1\%$ (assuming a typical value for optical cross-talk of 0.08~p.e.).

\begin{table}[htbp]
    \centering
    \caption{Results of the systematic study of the calibration parameters for a dark count rate of 3.08~MHz}
    \begin{tabular}{l|c|c|c|c|c}
      & $G$ & $\sigma_e$ & $\sigma_s$ & $\mu_{\textrm{XT}}$ & $\mu$ \\
        \hline
      Relative resolution [\%] & $\leq 0.1$ & $1$ & $7$ & $9$ & $1$ \\
      Relative systematic error [\%] & $\leq 0.1$ & $1$ & $2$ & $4$ & $1$ \\
    \end{tabular}
    \label{tab:systematics}
\end{table}{}

The distributions of the calibration parameters can be used to identify outlier pixels. 
Pixels for which at least one of the six calibration parameter values shows a 5 standard deviation from the camera-averaged value are flagged. 
In figure~\ref{fig:glob_fit_res_map} the identified outlier pixels are drawn in black. It can be seen that two outlier pixels are identified for each calibration parameter. For these pixels the MPE spectrum cannot be obtained, because the corresponding AC LEDs on the CTS are damaged. 
Furthermore, among the 1296 pixels, none showed to be inoperable. A few pixels have a lower gain than the central value or poor photo-electron resolution. But these out-off-spec pixels have been characterized ensuring raw data to photo-electron conversion.

\begin{table}[htbp]
    \centering
    \caption{Summary of the fit results.}
    \begin{tabular}{l |l | c | c | c}
        Parameters & Units & Mean & Std & Spread [\%]\\
        \hline
        Gain & [LSB/pe] & 20.47 & 0.54 & 2.6 \\
        $\mu_{\textrm{XT}}$ & [p.e.] & 0.08 & 0.01 & 12.5 \\
        $\sigma_s$ & [LSB] & 1.78 & 0.69 & 38.8 \\
        $\sigma_e$ & [LSB] & 3.37 & 0.40 & 11.81 \\
        Baseline residual & [LSB] & 1.17 & 0.68 & 58.12 \\ 
        Dark count rate & [MHz]                   & 3.08 & 0.67 & 21.7 \\
        $T$ & [$^\circ$C]  & 26.45 & 0.82 & 3.1 \\
    \end{tabular}
    \label{tab:glob_fit_res}
\end{table}

\paragraph{Gain equalization}\label{sec:gain_eq}

To reach a uniform signal response of the camera, each one of the 1296 read-out channel are equalized in gain. 
Each SiPM is provided with its operating voltage and a compensation loop on the slow control boards is implemented to cope with breakdown voltage changes with temperature, as described in~\cite{SST-1M-front-end}. 
To further improve signal uniformity the DigiCam allows to adjust the FADC gain by steps of $0.01\%$ for each readout channel within $\pm 5\%$ (see \cite{SST-1M-camera,SST-1M-Digicam-SPIE14} and \cite{SST-1M-Digicam-ICRC15} for more details) .

A gain uniformity of $2.6\%$ is achieved after one round of equalization as can be seen in figure~\ref{fig:glob_fit_res}. The gain is measured as described in section \ref{sec:mpe}.
The updated gains are equalized to the previously measured camera average gain. 
The uniformity can be further improved with many iterations, but 
a larger data sample is needed to achieve a precision below the percent level. 
The gain correction coefficients are loaded at each boot of the camera. 
This equalization process can also be regularly performed on-site, with the CTS mounted in front of the camera. 

DigiCam also permits to equalize the baseline of each FADC within $\pm 138$~LSB.

The method is similar to the one applied for the gain equalization. The baselines were measured in dark conditions with the SiPMs at their operational voltage over a collection of ten thousand waveforms of 50 samples each. % A baseline uniformity of $0.3\%$ post equalization is achieved. Baseline measurements and equalization are performed before each observation night with closed camera lid. The process itself takes a few minutes to complete. The dataset acquired before each observation run is also used for other monitoring purposes.

\paragraph{Pulse template\label{sec:pulse_template}}

As can be seen in section~\ref{sec:time-reco}, the pulse template $h(t)$ is used to reconstruct the timing information of the signal. It is also used for Monte Carlo simulations of the telescope as it describes the typical signal response to a single photo-electron signal. In addition, it provides a conversion factor between a single photo-electron amplitude $\bar{G}$ and single photo-electron integrated charge $G$, as can be seen from equation~\ref{eq:gain_conversion}. Moreover, to be able to extract charge and timing information the pulse template characterisation in the saturation region of the pre-amplifier and of the FADC is important.

The time delay scan, discussed in section~\ref{sec:acquisition}, is used to evaluate the pulse template. 
The flashes of the CTS are triggered by the camera and a delay is adjusted by steps of 77~ps to vary the position of the light pulse in the sampling window.
Given that each LED has a different response to a given DAC value and that the DACs control groups of neighboring pixels, the homogeneity of the light is not guaranteed with the CTS. Therefore, the timing scan is done by scanning the AC DAC level to ensure that most of the pixels are illuminated between 50 and 400 p.e. This light level range ensures a good signal-to-noise ratio and no saturation.

\begin{figure}[htbp]
\begin{center}
\includegraphics[width=0.42\textwidth]{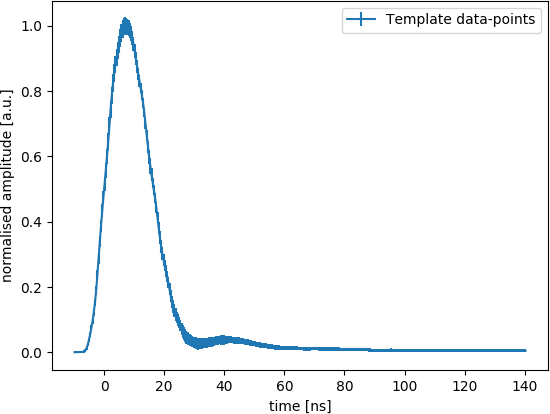}\hfill
\raisebox{-10pt}{\includegraphics[width=0.46\textwidth]{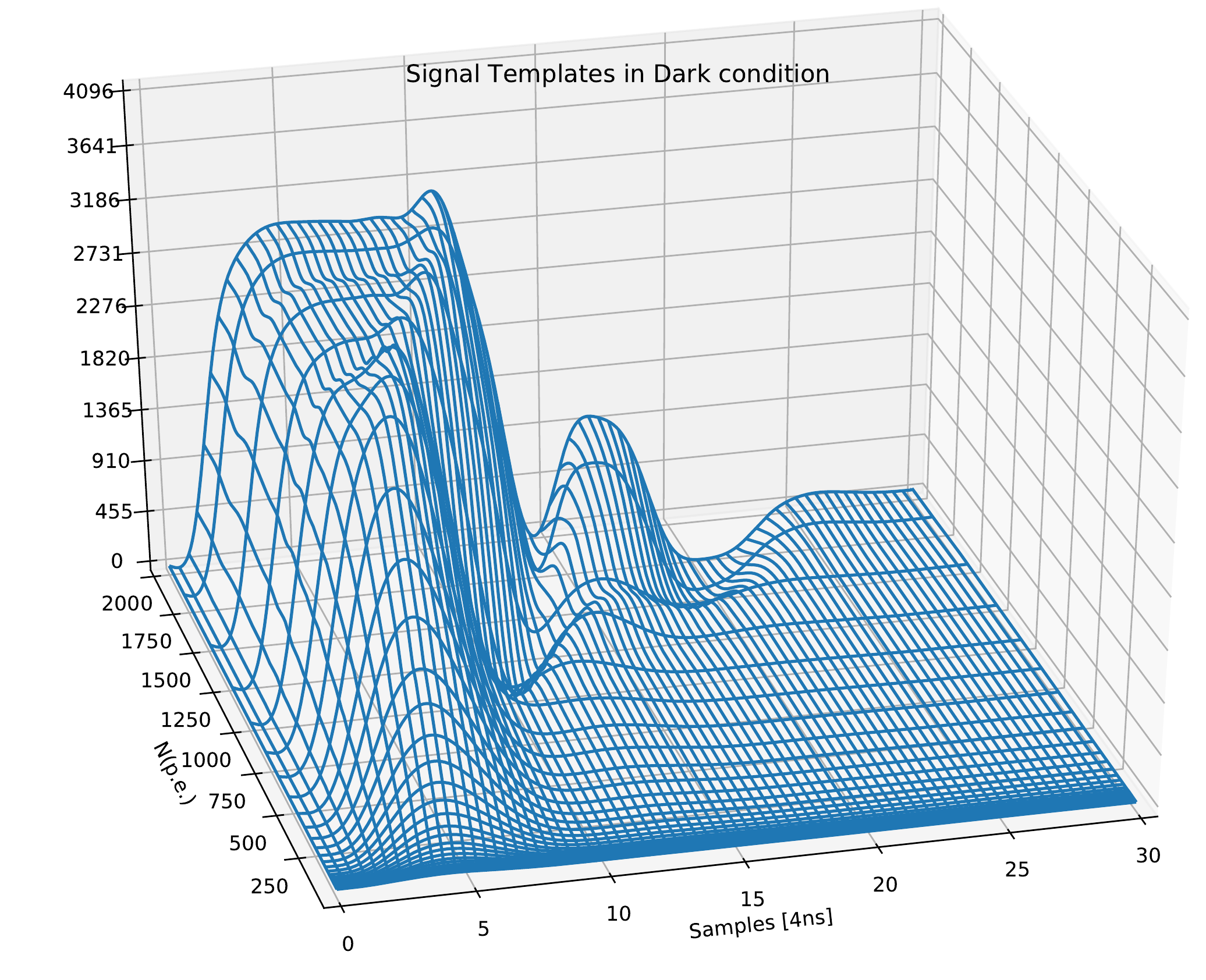}}
\caption{Left: Average normalized pulse template for the whole camera between 50 and 400 p.e. The error bar width corresponds to the standard deviation of the measurement for all pixels and all flashes. Right: Pulse template as a function of the number of photo-electrons.}\label{fig:template}
\end{center}
\end{figure}

For each camera pixel and each flash of light, a charge estimation is done summing up the FADC samples in a large enough window to fully contain the pulses for all LEDs, despite the offset in time between LEDs of the CTS. 
A large number of samples allows also to detect saturation when the pulse amplitude flattens, but the pulse width increases as shown in figure~\ref{fig:template}.

Pulses whose estimated charges are between 50 and 400 p.e are added to a single 2D histogram (combining all pixels). The histogram is binned as a function of the time of half maximum amplitude and as a function of the normalized amplitude (normalized by the estimated integral charge).
The template value and its uncertainty are taken as the mean and standard deviation of the normalized amplitude of each bin of time with significant number of entries. Figure~\ref{fig:template} shows the obtained average template.

\subsection{Optical element properties}

\begin{figure}
    \centering
    \includegraphics[width=0.45\textwidth]{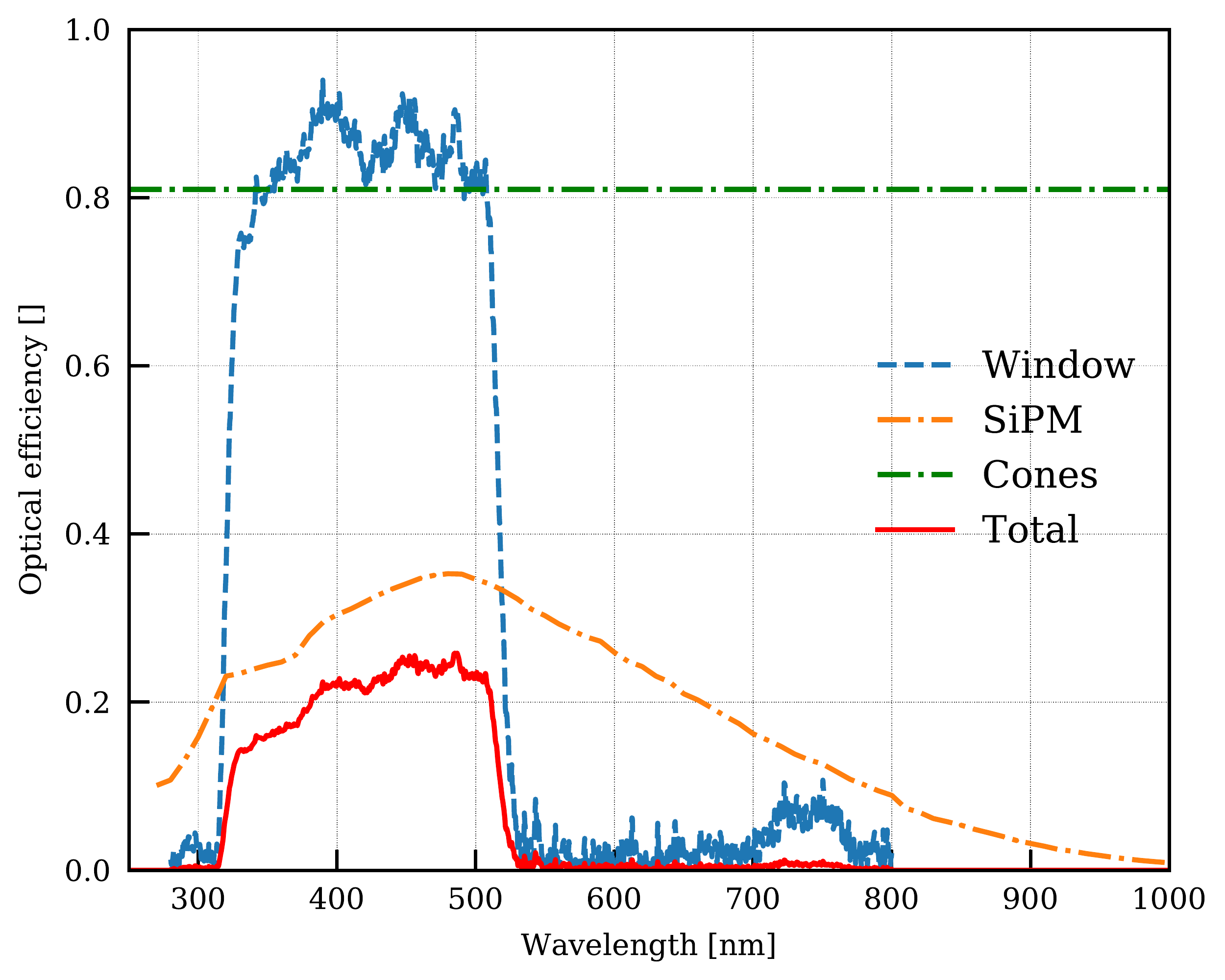}
    \includegraphics[width=0.45\textwidth]{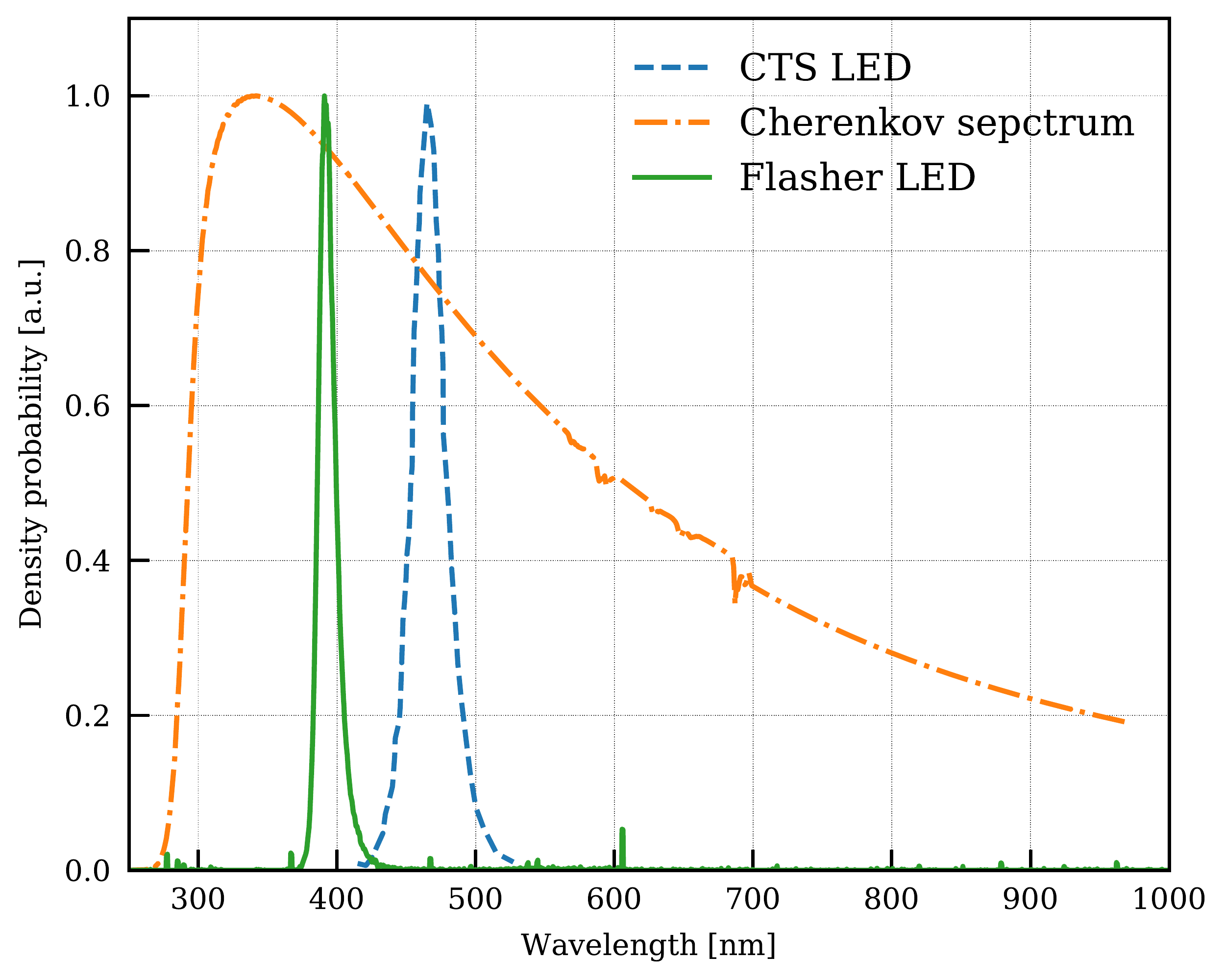}
    \caption{Left: Transmittance of the filtering window (blue), photo-detection efficiency of the SiPM (orange), the collection efficiency of the cones (green) and the overall efficiency (red) as a function of the wavelength. Right: Spectra of the calibration devices measured with a spectro-photometer (AvaSpec-ULS3648) and the CTA reference Cherenkov light spectrum in the atmosphere.}
    \label{fig:optical_efficiency}
\end{figure}

The efficiencies $\eta_i$ of the different optical components (namely the light funnels, the SiPMs and the entrance window) have been measured for the wavelengths in the range of the spectrum of Cherenkov emission in the atmosphere. The efficiencies as a function of the wavelength are shown in figure~\ref{fig:optical_efficiency} (left) together with the product of the efficiencies (in red). 
Their measurements provide the photo-electron to photon conversion factor, as illustrated in equation~\ref{eq:gamma_conversion}. 
The average efficiencies for all elements, convoluted with the Cherenkov spectrum over the wavelength range 300~nm to 550~nm (as defined in the CTA requirement B-TEL-1170), are provided in table~\ref{tab:optical_efficiency}. Also the total PDP efficiency, $\eta_{\textrm{total}} =  \eta_{\textrm{window}}\times\eta_{\textrm{cones}}\times \eta_{\textrm{SiPM}} = 17\%$, is indicated.

The average optical collection efficiency for the light concentrators below the cut-off angle corresponds to $0.88$ for the Cherenkov light spectrum. Taking into account a geometrical fill factor of $0.92$ the average collection efficiency of the cones reduces to $\eta_{\textrm{cones}} = 0.81$.
The verification of the cone performance cannot be done for each individual camera cone, being too time consuming.
For this, in the validation phase of the technique (described in~\cite{SST-1M-light-cones}), the uniformity of the quality over a batch of production has been verified allowing to qualify a batch just measuring few samples.
The photo-detection efficiency (PDE) for SiPM is very uniform across each silicon wafer and overall for SiPMs of the same type, being solid state device produced with well controlled and reproducible techniques over the same batch of production. In addition, the manufacturer has to comply with precise specification on the spread in breakdown voltage between the four channels of a single SiPM (< 0.3~V).
The PDE measurement done to characterize our sensor are described in detail in~\cite{SST-1M-SiPM}.

%Measuring the absolute optical efficiencies of a pixel is also a quite time consuming procedure therefore it has been done only for a limited number of pixels, which are then used as reference pixels for cross-validation on-site.

\begin{table}[hbpt]
    \centering
    \caption{Window transmittance $\eta_{\textrm{window}}$, cone collection efficiency $\eta_{\textrm{cones}}$, SiPM efficiency $\eta_{\textrm{SiPM}}$ and total efficiency $\eta_{\textrm{total}}$ for the Cherenkov light spectrum from 300~nm to 550~nm (as defined in the CTA requirement B-TEL-1170). In this range the CTA requires $\eta_{{\rm total}} \geq 20\%$).}\label{tab:optical_efficiency}
    \begin{tabular}{l | c }
         &  Cherenkov Spectrum (300-550~nm)  \\
        \hline
        $\eta_{\textrm{window}}$ & 0.68 \\
        $\eta_{\textrm{cones}}$ & 0.81 \\
        $\eta_{\textrm{SiPM}}$ & 0.29 \\
        \hline
        $\eta_{\textrm{total}}$ & 0.17 \\
    \end{tabular}
\end{table}

Table~\ref{tab:optical_efficiency} gives the average efficiencies, but as shown in figure~\ref{fig:wdw_scan} (right), the optical efficiency varies from pixel to pixel.
These differences need to be evaluated to properly reconstruct the number of photons in each pixel (flat-fielding). 
The flat-fielding correction is one of the main goals of the on-site calibrations (see section~\ref{sec:on-site}).
This operation can be done at the site using an external light source, mounted at the center of the mirror dish, which illuminates the camera with a given known photon intensity. 
The number of photo-electrons in each camera pixel is measured and corrected for the light in-homogeneity. This in-homogeneity was measured in the laboratory using a calibrated photo-diode (see section~\ref{sec:flasher}). Therefore, the relative differences in the number of reconstructed photo-electrons between the camera pixel gives a flat-fielding correction. 
The flat-fielding correction comprises the window transmittance, the cone collection efficiency and the SiPM PDE.

\subsubsection{Window transmittance}\label{sec:window}

\begin{figure}[htbp]
    \centering
    \includegraphics[height=150pt]{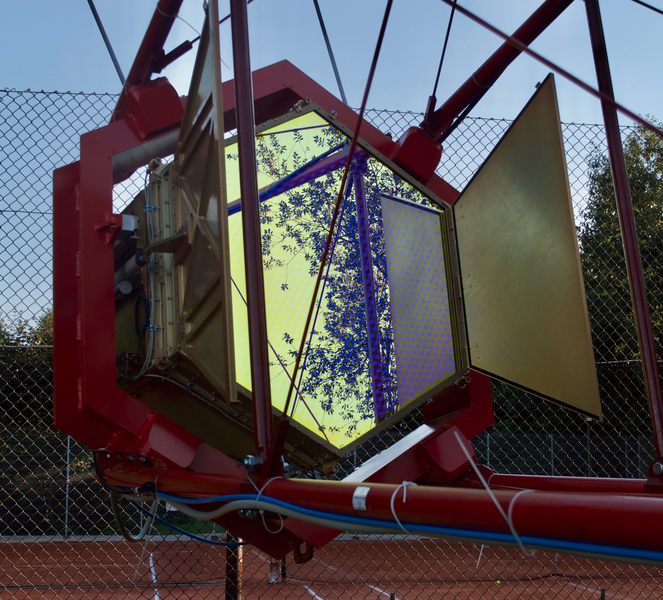}
    \includegraphics[height=150pt]{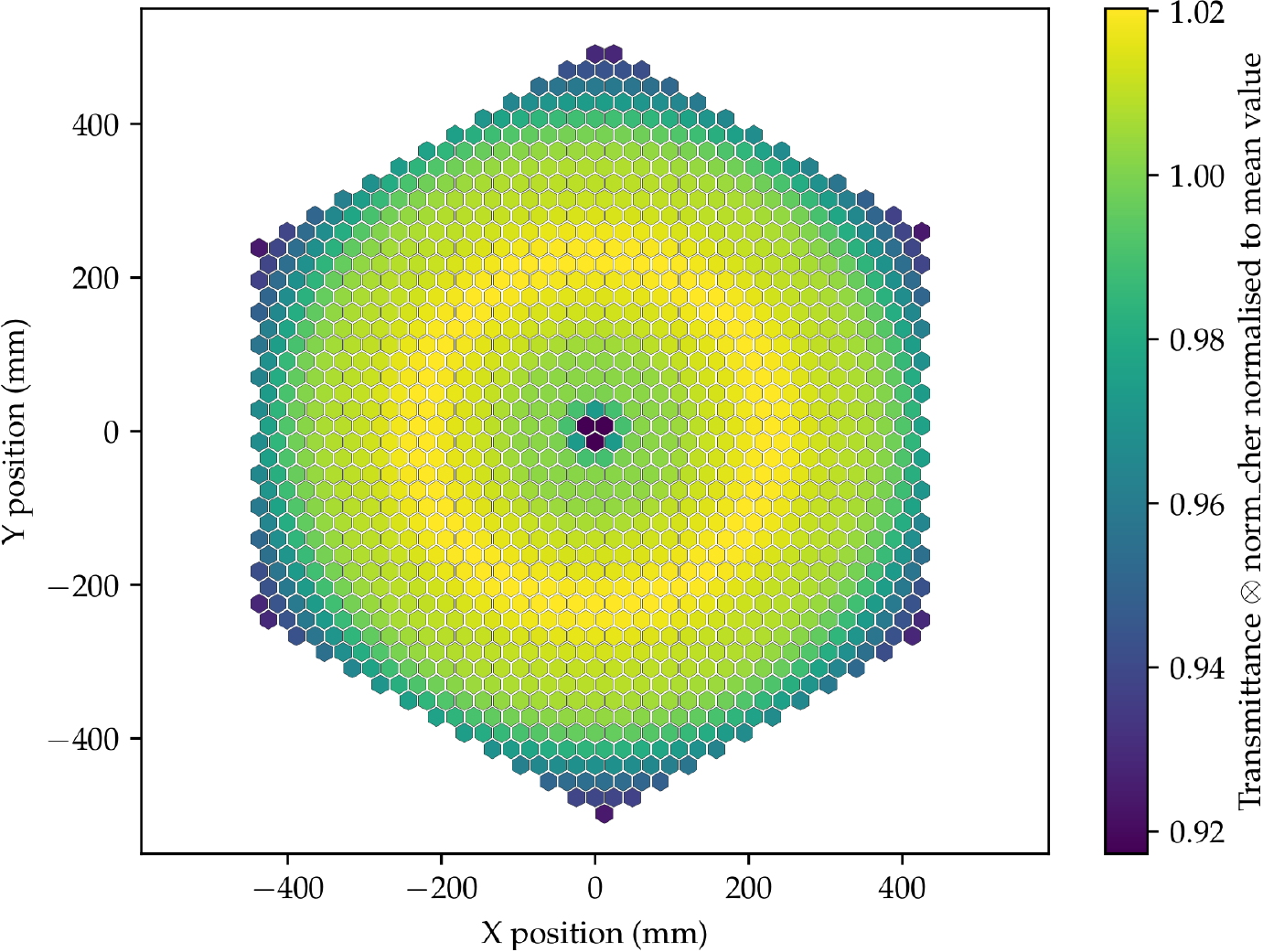}
    \caption{Left: The SST-1M camera equipped with its filtering window on the telescope structure in Krakow. The yellowish color of the reflected image is due to the window coating, which cuts off light beyond 540~nm, which is then reflected out of the camera. The filter is optimized for transmission in the blue. It also reduces Fresnel losses with an anti-reflective layer. Right: Relative efficiency of transmission for the Cherenkov light spectrum from 300~nm to 550~nm per camera pixel with respect to the average value over all pixels.}
    \label{fig:wdw_scan}
\end{figure}

The window coating is a complex process for such large optical elements as the window ($\simeq 0.81~{\rm m^2}$), given the scarce availability of large coating facilities. Additionally, it is extremely challenging to produce uniform coating layers over large surfaces, in order to guarantee uniform transmittance. 
To correct the images for disuniformities, the measured transmittance as a function of the wavelength and position across the window can be used. Figure~\ref{fig:wdw_scan} (left) shows the camera with the window installed on it.

An automatic test setup for the window quality check has been developed. The window is placed on a support moving along rails and connected to two linear motorized stages of $300$~mm travel range to scan the whole surface. A Deuterium-Halogen lamp AvaLight-DH-S from Avantes was used as light source, whose beam is coupled to a solarized UV-VIS $600~\mu$m diameter optical fiber equipped with a collimator which is placed on one side of the window. On the other side, a Labsphere 3P-GPS-053-SL integrating sphere, connected via an optical fiber to a spectro-photometer (AvaSpec-ULS3648 with a $25~\mu$m slit and a grating of $300$~lines/mm), is placed as light detector. Firstly the dark signal and reference signal without the filter window was captured. Then the filter window was automatically positioned between the collimator and the integrating sphere and the spectral transmittance was measured for a given window diagonal.
 
To optimize the coating uniformity, the window rotates in the coating chamber around its center and therefore, the coating profile has a radial symmetry.
Consequently, in principle it is sufficient to measure the transmittance only from one corner to the center, given the rotation symmetry. 
However, in this study, we have verified this hypothesis by measuring the 6 diagonals of the hexagonal window.

The window transmittance as a function of the wavelength is illustrated in figure~\ref{fig:optical_efficiency} (left) and it is consistent with the expected design specifications. Further details on the window transmittance can be found in~\cite{thesis-DeAngelis}. The multi-layer coating optical transmittance property depends on the light incidence angle. However, the maximal incidence angle of the light on the window is $25^\circ$ and the wavelength transmittance shift is negligible.  The window is effective in reducing the NSB as the expected rate, for a dark night, in each pixel is about 40~MHz (from $200$~nm to $1000$~nm) thanks to the low pass filter coating.
As matter of fact, with a simple window without any filter, even with a more UV transparent material as PMMA, the NSB would have been about 3.75 times bigger.
The non-uniformity among the six diagonals as a function of the distance to the center was measured to be lower than 2\%. 
Therefore, the average radial profile was used to generate the window transmittance map.
From this map, the relative per-pixel correction was derived and is shown in figure~\ref{fig:wdw_scan} (right). These corrections can be used to derive the per-pixel $\eta_{\textrm{SiPM}} \times$ $\eta_{\textrm{cones}}$ value from a measurement with uniform illumination or using muon ring images on-site (as described in~\cite{CTA-muon}). 

\section{Characterization of the camera performance}\label{sec:key_perf}

\subsection{Charge resolution}

The charge resolution plays a crucial role for the energy resolution of the telescope. 
The number of photons impinging on the camera and the density of photons at ground are directly related to the primary gamma-ray energy starting the electromagnetic shower. It is therefore fundamental to assess the error in reconstructing the number of photons and that this estimate is not affected by any systematic effect.
The charge resolution is ultimately limited by the nature of the source, which has Poisson fluctuations in the emitted number of photons. It also takes into account the resolution of the camera photo-sensors, pre-amplifying electronics and data acquisition chain. 

The data analysis performed here is similar to the one described in~\cite{SST-1M-camera}, but we now perform the measurement for each of the 1296 camera pixels and express results in photon units rather than photo-electrons. The results are also compared against the CTA requirements.

The charge resolution $CR$ is defined as the ratio between the variance and expected value of the reconstructed charge in units of photons:

\begin{equation}
    CR(N_{\gamma}) = \frac{Var(N_{\gamma})}{\mathbb{E}(N_{\gamma})}
\end{equation}

where, the variance and expected value are computed as standard deviation and mean of the acquired sample.
The charge resolution presented here takes into account: Poisson fluctuations of the LED source, electronic noise from the photo-detection plane, SiPM gain smearing, optical cross-talk, precision of the computed baseline, and non-linearity of the amplification chain. 

\begin{figure}[htbp]
\centering
\includegraphics[width=0.45\textwidth]{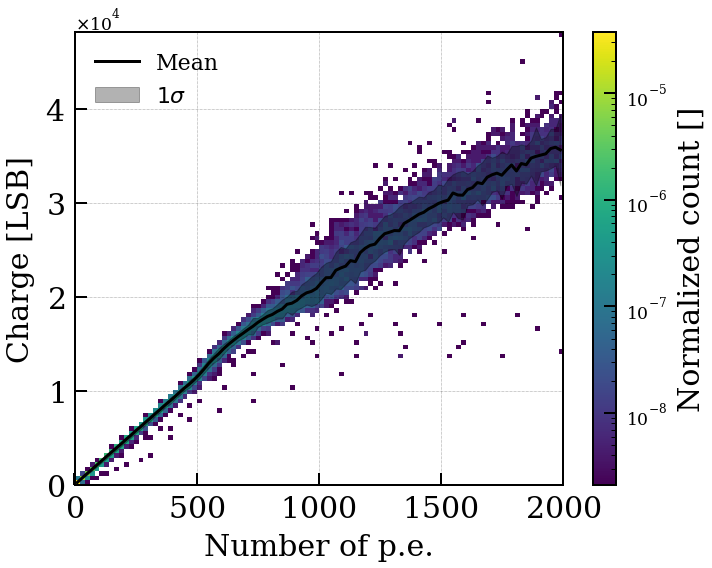}
\includegraphics[width=0.45\textwidth]{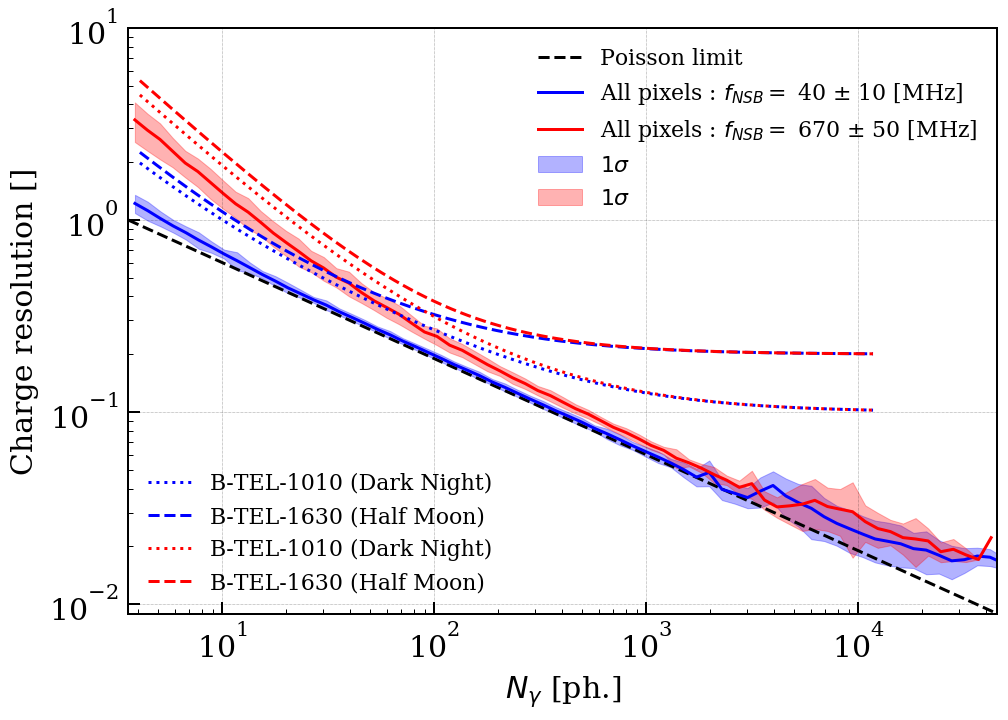}
\caption{Left: Charge linearity of all pixels. Right: Charge resolution in photons for all pixels and for two distinct NSB rates per pixel. The ranges of NSB rate correspond to dark sky conditions at 40~MHz, and a half Moon night at 670~MHz in Paranal, Chile. The CTA requirement curves, B-TEL-1010 and B-TEL-1630, for the small-sized telescopes of CTA are also drawn with dotted and dashed lines, which correspond to two data processing levels .}\label{fig:charge_resolution}
\end{figure}

In order to meet the requirements on the dynamic range (0-2000~p.e.) and on the charge resolution (see figure~\ref{fig:charge_resolution}) and in order to have a single amplifying channel per pixel, which is mandatory to keep the cost and the power consumption reasonable, the analogue stage has to saturate. The saturation of the response allows to reach the required charge resolution across the whole dynamic range, while the signal shape remains within the 12-bits range of the FADCs. This feature allows to retrieve as much information as possible from the digital pulse, which would be lost in the case of a digital saturation. 
For this reason, the measurements of the gain $G$ obtained from the MPE fit are reliable only in the linear regime. 
In figure~\ref{fig:charge_resolution} (left), the pulse integral as a function of the number of p.e. was extracted. 
As can be seen in figure \ref{fig:template}, the pulse shape is conserved up to saturation. 
A saturation threshold of 600~p.e. (respectively 3500~LSB) on average for all pixels is obtained. It is defined as a deviation from $>20\%$ of a linear extrapolation of the slope $G$, which is determined in section~\ref{sec:mpe}. 
To cope with the saturation of the readout chain the waveform integration window is extended, as the pulse width increases with increased light intensity.

The estimated number of photons sent by the LED is obtained from the number of photo-electrons $\mu_{ij}$ extracted from the MPE spectrum fit described in section \ref{sec:mpe}. 
The number of photons is then assessed by correcting for the efficiencies of the SiPMs, the filter window transmittance and the light concentrators as in equation~\ref{eq:gamma_conversion}. The effect of voltage drop from the NSB (see~\cite{SiPM-NSB}) is also taken into account in the correction of the number of photons.

The charge resolution, presented in figure~\ref{fig:charge_resolution} (right), has been measured with the AC/DC scan data. 
The solid lines represent the average resolution over the camera pixels and the contoured area represents its 1-$\sigma$ deviation. 
The theoretical Poisson limit (fluctuations of the light source) is shown as a black dashed line. 
The charge resolution is given for two distinct NSB rate levels, given as a rate of photo-electrons per pixel in MHz. The effects of the saturation can be seen at around four thousand photons but are well coped with as the resolution does not drastically worsen. 

\subsection{Time resolution}

The AC/DC scan data are also used to evaluate the time resolution of the camera pixels. 
The CTS LEDs were triggered by the camera in order to have the light pulse always in the same position within the sampling window.
The reconstructed time $t^\prime$ is obtained from the $\chi^2$ minimization as defined in equation~\ref{eq:chi2_timing}. It is calculated over the time window every $dt = 0.1$~ns (as smaller steps did not improve the results), for each pixel and each flash of light. 
The time with the lowest $\chi^2$ is chosen as the reconstructed time $t^\prime$ for the event. The time resolution for each pixel is obtained as the standard deviation of the sample.
Flashes of light whose measured amplitude are below 2.5 p.e. are not used in order not to affect the estimate of the time resolution with thermal noise and NSB.

\begin{figure}[htbp]
\begin{center}
\includegraphics[width=0.7\textwidth]{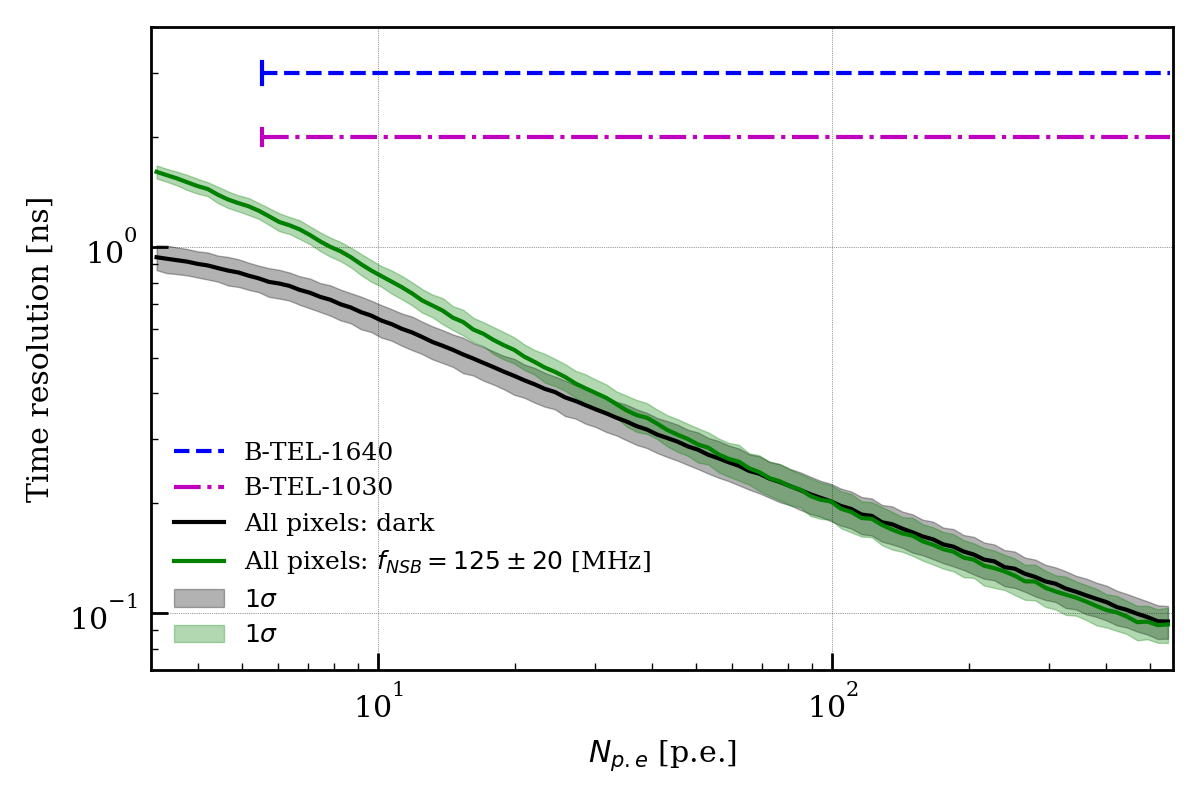} 
\end{center}
\caption{Time resolution of all pixels as a function of the number of photo-electrons in dark conditions (black) and with 125~MHz NSB (green). The time resolutions are compared to the CTA requirements as a reference for 125~MHz NSB and two data processing levels (B-TEL-1640 and B-TEL-1030). }\label{fig:time_resol}
\end{figure}

This measurement is repeated for several LEDs pulse amplitudes and the corresponding number of p.e. is derived using the LED calibrations (see figure~\ref{fig:leds}).
Figure~\ref{fig:time_resol} shows, for the whole camera, the evolution of the time resolution as function of the charge ($N_{\textrm{p.e.}}$) for 0 and 125~MHz NSB rate. 
As can be seen, the time resolution, without any NSB, is always below 1~ns and reaches 0.1~ns at 400~p.e. At 125~MHz of NSB, the resolution is mainly affected below 50 p.e. and goes above 1~ns only for pulse amplitudes below 7 p.e. (30 photons).

\section{On-site calibration strategy\label{sec:on-site}}

During the summer 2018, the SST-1M camera was mounted on the SST-1M telescope at IFJ PAN in Krak\'ow. This period allowed to commission the telescope and perform preliminary science observations. The site conditions are, however, not adequate to perform gamma-ray astronomy nor calibration tasks. In fact the telescope is located within the institution area and is exposed to many sources of background light: street lights; office lights; nearby airport; nearby national road. The NSB light rate per pixel was estimated to range between 0.5~GHz and 1.5~GHz depending on the cloudiness of the sky~\cite{SST1M-project-ICRC19}. When compared to the darkest nights, at the foreseen CTA southern site, this is at least 12.5 times higher. Additionally, the external light source was not yet mounted on the telescope structure. For these reasons most of the calibration tasks at the site could not be performed. In 2021, the telescope will be installed at the Ond\v{r}ejov Observatory in Czech Republic. Following this, the on-site calibration tasks, described in the following, will be performed there.

The goal of the on-site calibration activities is to monitor that the calibration parameters, listed in table~\ref{tab:glob_fit_res} and table~\ref{tab:optical_efficiency}, extracted off-site and used for the image reconstruction still apply when operating at the site. Moreover, it also allows to determine the flat-field coefficients. 
To do so, three event types can be used for each observation night, i.e. muon events, dark count events and external light source events.

Muon events (see section~\ref{sec:muon_events}) allow to monitor the stability over time of the optical throughput of the telescope,  providing an overall correction factor for telescope shadowing, mirror reflectivity, window transmittance, cone collection efficiency and sensor PDE.

The dark count events (see section~\ref{sec:dark_runs}) allow to reproduce data that were acquired in the laboratory and can, therefore, be directly compared to it. If a significant discrepancy between on-site and laboratory results is observed, the parameters in equation~\ref{eq:pe_conversion}, used to reconstruct the number of photo-electrons, have to be re-evaluated to ensure unbiased photo-electron reconstruction.

The external light source events (see section~\ref{sec:flasher_events}) are affected by the ambient background light (typically 40~MHz night-sky background for a moon-less night), which means that SiPM parameters cannot be extracted as done in section~\ref{sec:mpe}. However, provided that the light source has been thoroughly characterized, it allows the monitoring of the model presented in~\cite{SST-1M-voltage-drop} and the extraction of the flat-field coefficients to equalize the trigger among the pixels as presented in section~\ref{sec:trigger_flat_field}.

Additionally, as mentioned in section~\ref{sec:ls_cts}, the CTS can be used on-site for a re-calibration or to identify components that need replacement. The re-calibration can also be performed yearly independently of any corrective maintenance activities.

\subsection{Muon events}\label{sec:muon_events}

Muon events, which produce ring images or arc images in the camera plane, are expected to be triggered with the same trigger configuration as the one used for the atmospheric showers. Dedicated muon triggers, more adapted to the ring-like structure, are possible to tag the muons on-line without requiring any data processing.
The overall optical efficiency of the telescope can be assessed by comparing the expected number of photons within a muon image (assessed from Monte Carlo simulations) to the reconstructed number of photo-electrons within the muon image. The method is widely used in the current generation of IACTs as can be found in~\cite{MAGIC-muon, HESS-muon, VERITAS-muon}. A dedicated study for the SST-1M telescope was carried on and is presented in~\cite{CTA-muon}.

\subsection{Dark count events}\label{sec:dark_runs}

\begin{figure}
    \centering
    \includegraphics[width=0.49\textwidth]{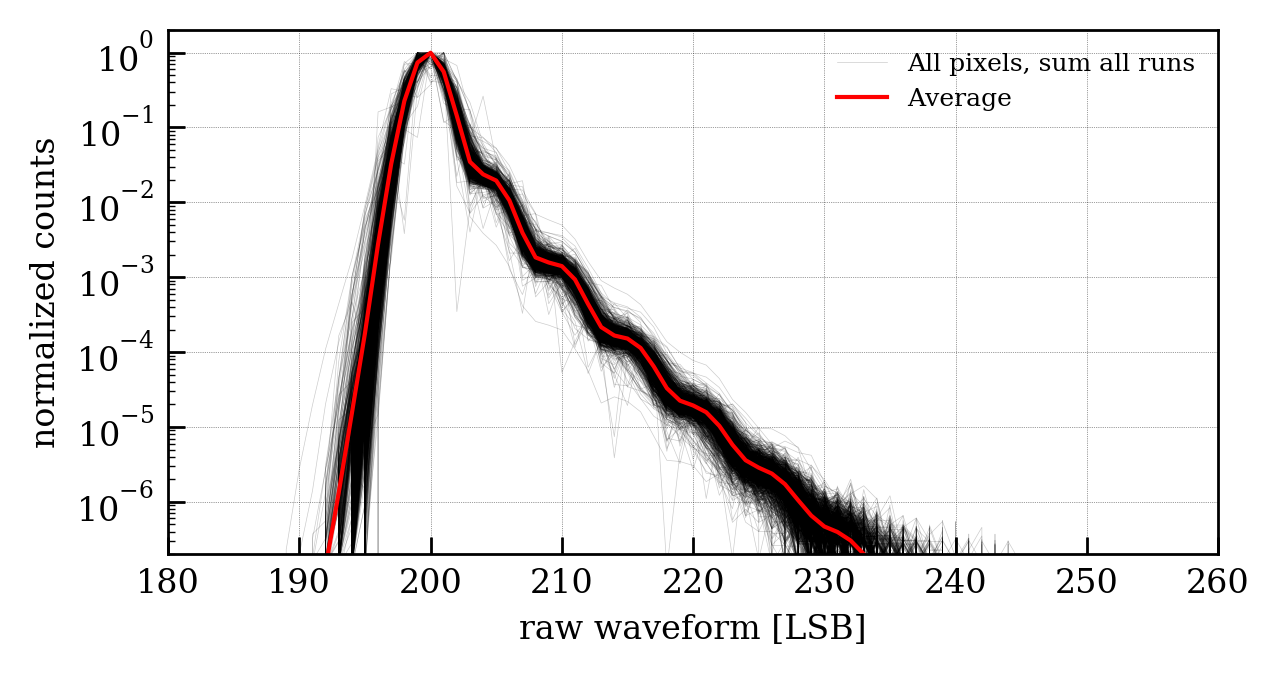} \hfill
    \includegraphics[width=0.49\textwidth]{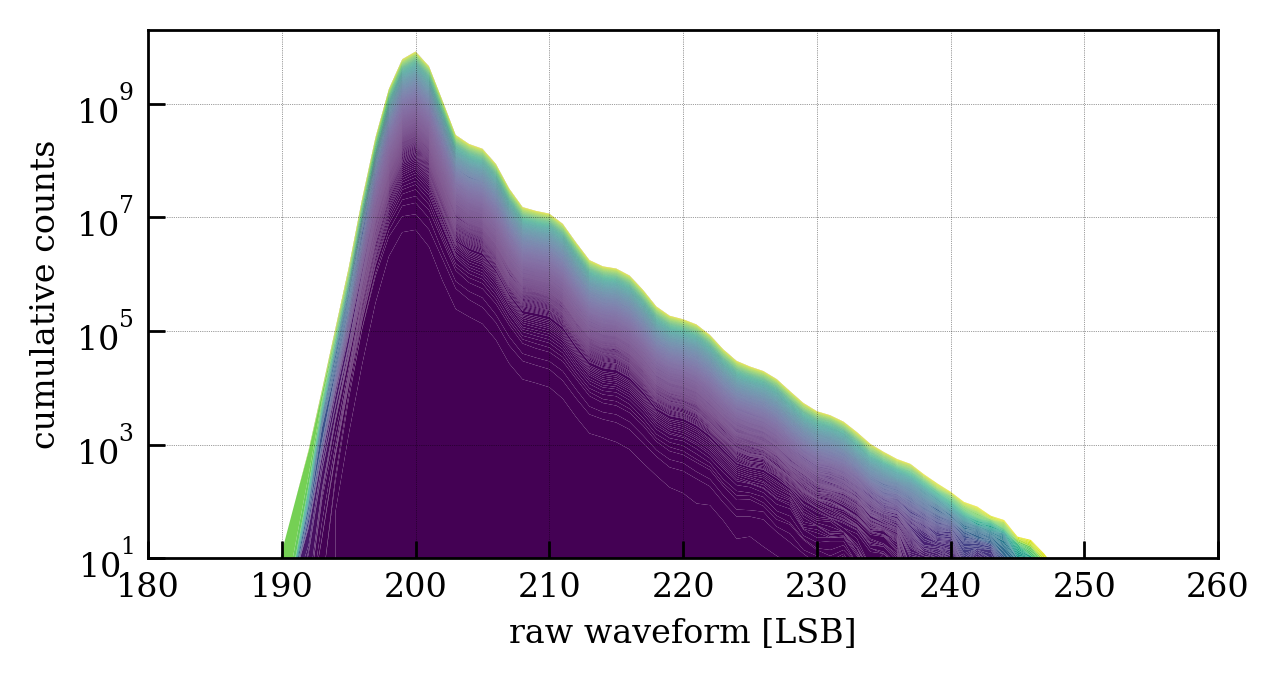}
    \caption{Left: Normalized distribution of the projection of the raw waveform in LSB for all the pixels and one run. Right: Merged distribution of all pixels and all nights.}
    \label{fig:dark_spe}
\end{figure}

The dark counts events are acquired at a fixed rate of 1~kHz using the internal clock of the digital readout. Dedicated dark runs are envisaged each night before and after sky observation, while the lid is closed. 
During each dark run ten thousand waveforms for each pixel are registered. The charge spectra can be used to evaluate the SiPM parameters. Even though the SiPM parameters cannot be measured with the same precision as presented in section~\ref{sec:mpe}, it is a useful procedure to track their evolution with temperature. 
It is worth noting that the dark count events also probe the camera component aging, such as the front-end and digital electronics aging. Figure~\ref{fig:dark_spe} shows the projection of the raw waveform in LSB for each camera pixel. 
Up to 6 p.e., the sensor parameters  are easily measurable on a per night basis and can be extracted with satisfactory precision. Additionally, merging all the dark runs and/or merging all the pixels allows building up the necessary statistics to correct the optical cross-talk model, as shown in~\cite{FACT-calib}.

The dark count runs are also used to measure the baseline in the absence of NSB. This is used to determine the baseline shift per pixel induced by the NSB. The baseline shift is used to correct the pixel gain, photo-detection efficiency and crosstalk as described in~\cite{SiPM-NSB}. The baseline shift is also used to evaluate the NSB.% as shown in~\cite{SST1M-project-ICRC19}.

\subsection{External light source events}\label{sec:flasher_events}

The external light source, which will be referred to as flasher in the following sections, is mounted at the focal distance of the telescope, in the middle of the mirror dish but not exactly on the optical axis as this position is occupied by the lid CCD camera used for the telescope pointing. Short light flashes (few nanoseconds) will be emitted during the science data taking at a rate of $100$~Hz. This rate is low enough not to affect the physics data taking. Events are acquired using the flasher which acts as a trigger source. Upon reception of a trigger signal emitted by the light source, the readout of the entire camera is started and the event is tagged with a specific trigger type.

\subsubsection{Characterization of the external light source}\label{sec:flasher}

The flasher (LUMP Calibration box), has been designed and built by the astroparticle group at the University of Montpellier for the CTA medium-sized telescope (MST)~\cite{CTA-MST} and having in mind the CTA requirements. 
It consists of an IP65 box containing an array of 13 LEDs whose peak wavelength is 390~mn (see figure~\ref{fig:optical_efficiency}-right) to resemble the Cherenkov spectrum, followed by a diffuser which is transparent to wavelength from 350 to 800~nm and whose diffusion angle is equal to 10$^\circ$. The FWHM of the pulsed light is between 4 to 5~ns, similar to Cherenkov flashes. In addition, the user can turn on and off individual LEDs, vary their intensity and adjust the frequency of the flashes from 100~Hz to 10~kHz.

\paragraph{Temperature characteristics}

\begin{figure}
    \centering
    \includegraphics[width=0.47\textwidth]{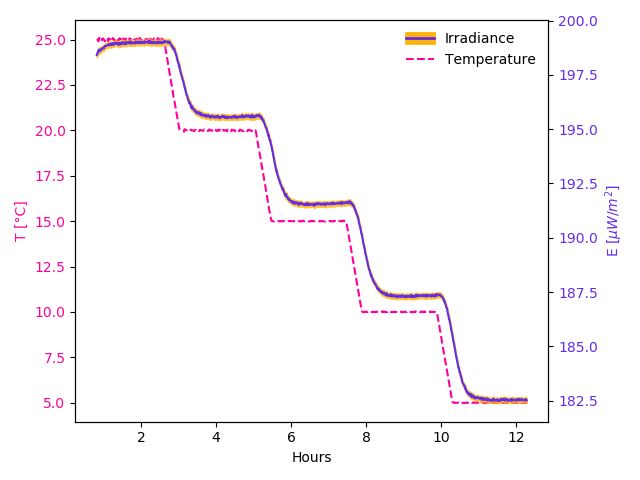} \hfill
    \includegraphics[width=0.52\textwidth]{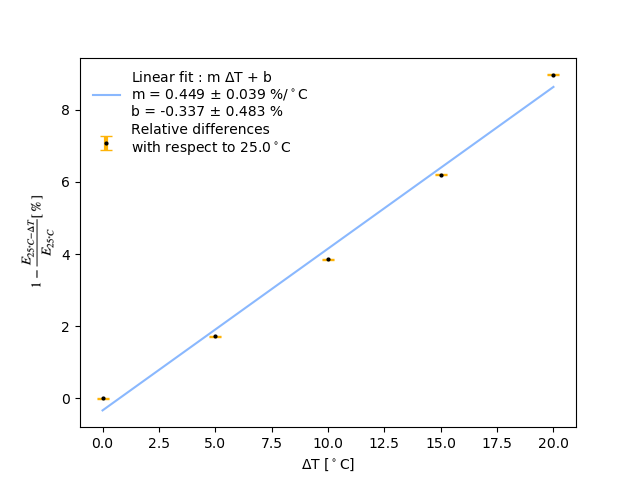}
    \caption{Left: Irradiance drops with temperature, purple solid line, with a step-like behavior at every stable temperature range, as it follows the temperature curve described by the climatic chamber, dashed pink line. Right: Projected relative irradiance variation as a function of temperature variation. Here variations are done with respect to room temperature of $25~{\rm ^\circ C}$.}
    \label{fig:temperature_characteristics}
\end{figure}

The flasher behavior versus temperature changes has been studied in a climatic chamber with a 3$\times$3~cm$^2$ photo-diode (S3584-08 designed by Hamamatsu).
To have control over the photo-diode as well as the flasher temperatures, both are placed within the climatic chamber. 
The temperature range is varied from $25~{\rm ^\circ C}$ to $5~{\rm ^\circ C}$ in steps of $5~{\rm ^\circ C}$ while the flasher illuminates continuously the photo-diode. 
The temperature of the climatic chamber as a function of time is shown in figure~\ref{fig:temperature_characteristics} (left) together with the measured irradiance on the photo-diode. The relative variation of the irradiance for each temperature difference with respect to 25~${\rm ^\circ C}$ is shown in figure~\ref{fig:temperature_characteristics}-right. 
The slope gives the value of the variation per unit of temperature degree equal to $0.449 \pm 0.039$~\% per ${\rm ^{\circ}C}$. According to the manufacturer, the photo-diode efficiency drops by 0.1~$~{\rm\%/ ^\circ C}$ at a wavelength of 390~nm. 
Therefore, the contribution alone from the flasher to this variation is $0.35 \pm 0.04$~\% per $^\circ {\rm C}$. For instance, at Paranal observatory, the temperature of a typical night drops at a rate of $0.4~{\rm ^\circ C/h}$~\cite{astrometeo}. Thus for a 10 hours observation period, the light yield of the flasher would be decreased by $1.4\pm0.16\%$ with respect to the light yield at the beginning of the observation.

\paragraph{Light stability over time}

As the flasher will be operated over the full observation period, it is crucial to characterize its light stability over time regardless of the temperature variations. 
The stability over time was assessed with a photo-diode mounted at a distance of 1~m from the flasher. The same photo-diode as for the temperature characteristics measurement has been used. The flasher illuminates at 5~kHz the photo-diode in a dark box whose temperature was kept at $21 \pm 1~^{\circ}{\rm C}$. 

\begin{figure}[htbp]
    \centering
    \includegraphics[width=0.5\textwidth]{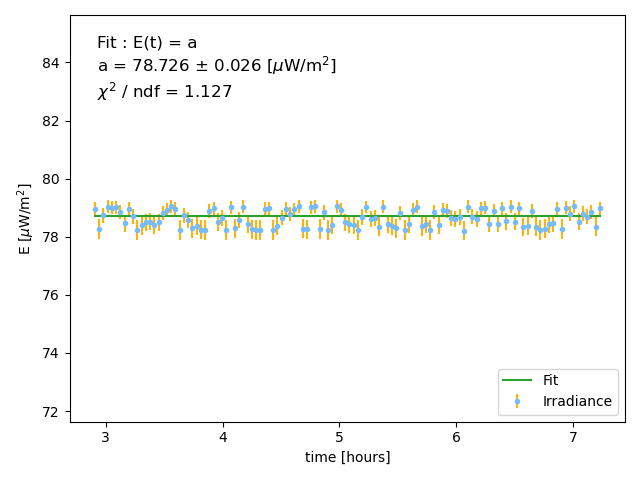}
    \includegraphics[width=0.45\textwidth]{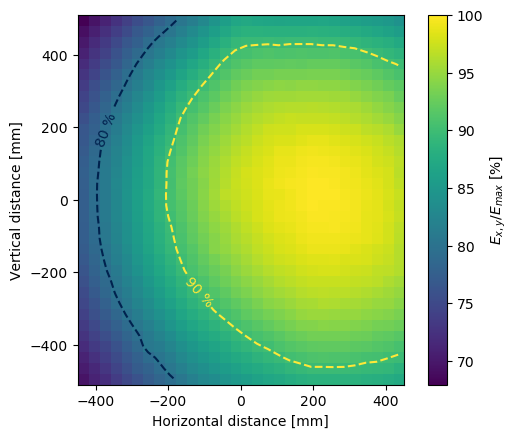} 
    \caption{Left: Flasher irradiance as function of time. A constant function is fitted (green). Right: Relative irradiance profile with respect to the maximum irradiance.}
    \label{fig:flasher_stability_over_night}
\end{figure}

This measurement shows a stable irradiance over 4 hours as shown in figure~\ref{fig:flasher_stability_over_night} (left). The data was better-fitted with a constant function rather than with a first order polynomial. In addition, the fluctuations of the light in this time period correspond to 0.4\% of the light yield. The error bars and the temperature fluctuations are compatible with the irradiance fluctuations.

\paragraph{Spatial light profile}

\begin{figure}
    \centering
    \includegraphics[width=0.9\textwidth]{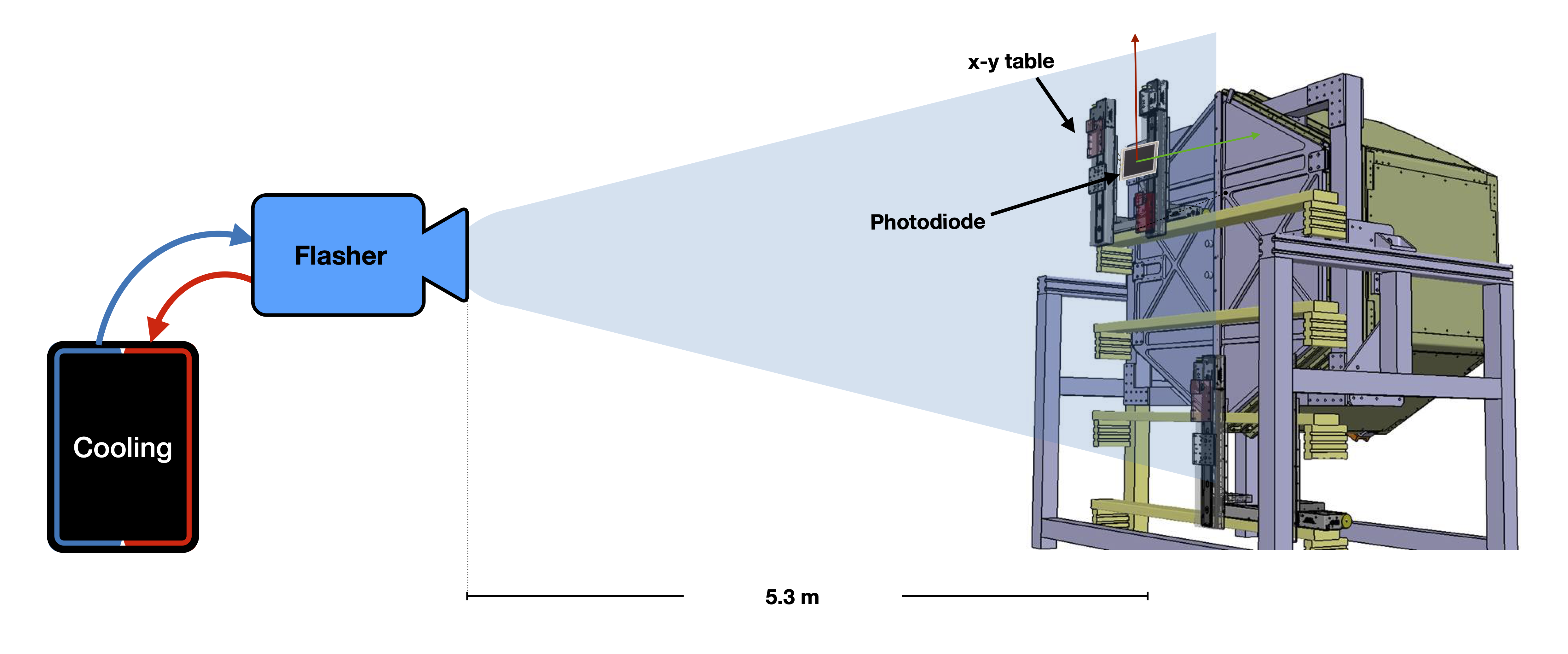}
    \caption{Scheme of the experimental setup used for the measurement of the flasher spatial light profile. The flasher is placed 5.3~m away from an x-y table scanning the camera surface. The flasher is cooled to keep its temperature constant. A photo-diode measures the irradiance.}
    \label{fig:scheme_flasher_setup}
\end{figure}

The spatial light profile must be known to be able to extract the relative optical efficiency of all the camera pixels. 
To measure this spatial light profile, mechanical supports have been built in front of the camera to host an x-y table and scan a total surface of 1.08~m$^2$, bigger than the PDP surface. The light profile is scanned using the same photo-diode as in the previous measurements. The flasher device is settled at a distance of 5.6~m away from the PDP corresponding to the focal length of the telescope (see figure~\ref{fig:scheme_flasher_setup}). However, the sensor is positioned at 5.3~m away from the flasher device, due to the mechanical constraints. The alignment with respect to the center of the camera is not trivial to obtain manually as seen by the presence of an offset between the center of the PDP and the maximum intensity of the light pattern (see figure~\ref{fig:flasher_stability_over_night} (right)).

As shown in figure~\ref{fig:flasher_stability_over_night} (right), if the flasher is well aligned with respect to the camera center, the irradiance of the flasher is relatively constant as the minimum irradiance at the boundaries is about 90\% of the maximal irradiance.

\subsubsection{Monitoring of optical efficiencies with muon and flasher events}

The degradation of the mirrors performance is expected due to dust deposition or degradation induced by environmental condition, e.g. acid rain, bird feces, etc.
In particular, dust sedimentation has the tendency to increase diffuse reflectivity~\cite{AUGER-mirror} but can be recovered by cleaning the mirrors regularly. From the experience of current generation of IACTs, we can expect a decrease of specular reflectivity of about 2\% per year due to: dust deposition, mechanical damage of the upper protective layer and oxidation of the aluminium coating.
The window is also exposed to an open environment and damages on its surface are expected as well. However it is much less exposed as it is sheltered by the lid when not observing. Concerning the cones, we expect minimal degradation of their reflecting coating with time, as they are in the sealed environment of the camera. The same applies to the SiPM sensors, which are not subject to significant aging even when exposed to high light level. 

The total optical efficiency to the muon event $\eta^{\textrm{muon}}$ and to the flasher events $\eta^{\textrm{flasher}}$ are probing different optical elements as can be seen from equations \ref{eq:opti_effi_cher} and \ref{eq:opti_effi_flasher}. 

\begin{eqnarray}
        \label{eq:opti_effi_cher}
    \eta^{\textrm{muon}} &=& \eta_{\textrm{shadowing}} \times \eta_{\textrm{mirror}} \times\eta_{\textrm{window}} \times \eta_{\textrm{cones}} \times \eta_{\textrm{SiPM}} \\
        \label{eq:opti_effi_flasher}
    \eta^{\textrm{flasher}} &=& \eta_{\textrm{window}}^{\prime} \times \eta_{\textrm{cones}}^{\prime} \times \eta_{\textrm{SiPM}}^{\prime}
\end{eqnarray}

The difference between $\eta$ and $\eta^{\prime}$ comes from the spectrum of the light source: the Cherenkov spectrum for the muons and the LED spectrum for the flasher (see figure~\ref{fig:optical_efficiency} right).

An efficiency drop cannot be attributed to a given element as only the product is measured.
Using both muon and flasher events, the contribution from the mirrors from the camera can be disentangled where the wavelengths of the Cherenkov and flasher spectra overlap. For instance a relative drop of the optical efficiency measured with muon events indicates a degradation of the mirrors reflectance or/and of the window transmittance. In particular, if no relative drop of optical efficiency is observed with the flasher events, the drop can only be attributed to the mirrors. As both muon and flasher events are acquired during observations, this measurement can be performed on a per night basis.

As the SiPM response depends on the voltage drop, as described in section~\ref{sec:photon-reco}, the $\eta_{\textrm{SiPM}}$ has to be corrected according to the level of NSB. This effect is assumed to be corrected in equations \ref{eq:opti_effi_cher} and \ref{eq:opti_effi_flasher}, meaning that the drop in photo-detection efficiency, gain and optical cross-talk are accounted in the efficiency drop.

\subsubsection{Flat-fielding}\label{sec:trigger_flat_field}

To get a uniform trigger rate across the camera field of view, the SiPM bias voltage could be tuned individually. 
While achieving the desired goal, this would complicate the photon reconstruction as the parameters of the SiPMs reported in section~\ref{sec:off-site} are given for a nominal bias voltage. 
The same goal can be achieved with a different strategy, which is applying different trigger thresholds in different camera regions, called clusters.

For the standard triggering scheme (see~\cite{SST-1M-calibration-ICRC17} for more information on the trigger), the camera is divided into 432 clusters. Each cluster (except for the ones located at the camera border) comprises 7 neighboring triplets, where each triplets is composed by 3 pixels, for a total of 21 pixels (one pixel belongs to multiple clusters). The signal of each pixel is baseline subtracted (the baseline is computed as described in equation~\ref{eq:baseline_comp} by the DigiCam FPGAs). The signal of each triplet (sum of baseline subtracted triplet pixel signals) is clipped to form a 8 bit-integer and added to the one of the 6 other triplets. The resulting signal sum is compared to a threshold which can be adjusted for each cluster independently and dynamically. This last point is critical to be able to flat-field the trigger response. A global threshold $T$ is determined from validated Monte Carlo simulation for a given NSB level. Then the individual cluster thresholds $T_{\textrm{cluster}}^j$ are applied accounting for the possible non-uniformities of the pixel responses, i.e. the flat-field coefficients. The flasher events are used to compute these coefficients. The coefficient $F_i$ for the pixel $i$ is the ratio between the pulse amplitude $A_i$ measured in this pixel corrected for the light profile and the camera average pulse amplitude:
\begin{equation}
    F_{i} = \frac{A_{i}}{\left<A\right>}
\end{equation}

These coefficients not only account for different optical efficiencies between pixels but also for different electronic responses, including the effect of the voltage drop.
For each cluster $j$, the threshold is computed as in equation~\ref{eq:trigger_with_ff}.
\begin{equation}
    T_{\textrm{cluster}}^j = \frac{{T}}{21}\sum_{i=1}^{21} F_{i}.
    \label{eq:trigger_with_ff}
\end{equation}

\section{Conclusion and outlook}

We have presented the method to reconstruct the number and arrival times of photons for a gamma-ray telescope camera. The methods to extract the relevant calibration parameters have been presented and illustrated using the SST-1M camera prototype with its calibration devices. Moreover, we have applied the model of SiPM optical cross-talk described in~\cite{SiPM-Vinogradov} and have showed that it can be reliably used with a large number of sensors to extract the photo-sensors properties. It has been applied to various light levels (up to 200 p.e., i.e. well below the sensor saturation) reducing the number of free parameters, while increasing the sample size allowing to achieve competitive precision. Furthermore, we have calculated the systematic effects induced by dark count rate on the measurement and found that it is negligible. Since the optical cross-talk affects mainly neighboring cells~\cite{cross-talk-sipm} (the number of which is limited to 8) and the generalized Poisson model does not take into account this fact, we might expect systematic problems for higher cross-talk probability than the one reported for our sensor. The photo-sensor typical properties have to be injected in the Monte-Carlo to enable reliable predictions of the camera response. 
%A dedicated publication will focus on the SST-1M telescope sensitivities.

The full calibration procedure is automatized thanks to the integration of the various sub-systems being controlled by the telescope control software. The data have been processed with the open-source pipeline of the SST-1M telescope~\textit{digicampipe} that relies on widely-used python libraries. 

The results of the off-site calibration show reliable performances for gamma-ray astronomy when compared to the requirements of the next-generation gamma-ray observatory. In particular: A gain uniformity of the entire readout chain of 2.6\% was observed. By adjusting the trigger delays a precise determination of the pulse template in steps of 77~ps is obtained even with a 4~ns FADC sampling period. The template has been characterized above the linear range allowing to use its features for the charge and arrival time reconstruction for saturated pulses. The efficiencies of the optical elements have been presented. In this paper, the window transmittance as a function of wavelength was determined on the full surface. The measurement showed irregularities of the coating on a surface as large as 0.81~m$^2$. However, the irregularities have been characterized and are comparable in all radial directions. Thus they can be accounted for in the photo-electron to photon conversion. Moreover, we show that even on large Cherenkov cameras a filtering window can be used to reduce night-sky background light without losing too much of the Cherenkov light. The overall optical efficiency of the PDP of $\eta_{\textrm{total}} = 17\%$ was found for the Cherenkov light spectrum. To increase the PDP optical efficiency above 20\% (in order to comply with the corresponding CTA requirement), a second window was designed extending the transmittance bandwidth to lower wavelength. Additionally, replacing the S10943(X) SiPMs with the latest low cross-talk technology from Hamamatsu, S13360(X), could be considered but this would require redesign of the front-end electronics. The reported charge and time resolution show excellent performances and satisfy highly demanding requirements, for instance those of CTA. 

The results of the on-site calibration show that the SiPM degradation can be monitored by performing dark runs on a per-night basis before astronomical observations. We characterized the flasher that will be used as a flat-field device while the telescope is observing the sky. The flasher will ensure a uniform trigger response of the camera over the night even in the presence of a highly fluctuating night-sky background level. In the future, we would like to show that the multiple photo-electron spectrum analysis performed off-site can be repeated on-site with the flasher. This would require the inclusion of a NSB rate parameter in the likelihood similarly as done in~\cite{SiPM-spectra}. However, we expect it could only be performed during the darkest nights (with a night-sky background less than $60$~MHz). Following this, we would like to provide an absolute calibration of the flasher allowing for absolute monitoring of the optical efficiencies. 

\section*{Acknowledgements}
This work was supported by the grant Nr. DIR/WK/2017/12 from the Polish Ministry of Science and Higher Education. We greatly acknowledge financial support form the Swiss State Secretariat for Education Research and Innovation SERI. The work is supported by the projects of Ministry of Education, Youth and Sports of the Czech Republic: MEYS LM2015046, LM2018105, LTT17006, EU/MEYS CZ.02.1.01/0.0/0.0/16\_013/0001403 and CZ.02.1.01/0.0/0.0/18\_046/0016007 Czech Republic. This work was conducted in the context of the CTA SST-1M Project.

\bibliographystyle{JHEP} 
\bibliography{paper.bib} 

\end{document}